\documentstyle[amssym,psfig]{mn}

\newif\ifAMStwofonts

\newcommand{\dapp}{$\approx$}
\newcommand{\app}{$\sim$}
\newcommand{\tento}[1]{$10^{#1}$}
\newcommand{\vzs}{$\times$}
\newcommand{\lb}{$\lambda$}
\newcommand{\mm}{$\pm$}
\newcommand{\ion}[2]{\hbox{#1\,{\sc {#2}}}}
\newcommand{\msol}{$M_\odot$}
\newcommand{\ebv}{\hbox{$E(B\!-\!V)$}}
\newcommand{\hbc}{H$_\circ$}
\newcommand{\av}{A$_V$}

\newcommand{\jmh}{\hbox{$(J\!-\!H)$}}
\newcommand{\hmk}{\hbox{$(H\!-\!K)$}}
\newcommand{\kms}{km\,s$^{-1}$}
\newcommand{\ergs}{ergs\,s$^{-1}$}
\newcommand{\ergcs}{ergs\,cm$^{-2}$\,s$^{-1}$}

\newcommand{\mic}{$\umu$m}
\newcommand{\brgama}{Br$\gamma$}
\newcommand{\padelta}{Pa$\delta$}
\newcommand{\pagama}{Pa$\gamma$}
\newcommand{\pabeta}{Pa$\beta$}
\newcommand{\palfa}{Pa$\alpha$}
\newcommand{\halfa}{H$\alpha$}
\newcommand{\hbeta}{H$\beta$}
\newcommand{\bsiiia}{[\ion{S}{iii}]\,\lb9532}
\newcommand{\bsiiib}{[\ion{S}{iii}]\,\lb9068}
\newcommand{\bhen}{\ion{He}{i}\,1.083\mic}
\newcommand{\bfeiia}{[\ion{Fe}{ii}]\,1.26\mic}
\newcommand{\bfeiib}{[\ion{Fe}{ii}]\,1.64\mic}
\newcommand{\hh}{H$_2$}
\newcommand{\hhl}{H$_2$ $\nu$=1$-$0 S(1)}
\newcommand{\bhepg}{\ion{He}{i}$+$\pagama}
\newcommand{\bhepd}{\ion{He}{ii}$+$Pa$\delta$}
\newcommand{\feii}{\ion{Fe}{ii}}
\newcommand{\niit}{[\ion{N}{ii}]\,\lb}
\def\aj{AJ}
\def\araa{ARA\&A}             
\def\apj{ApJ}                 
\def\apjs{ApJS}               
\def\aap{A\&A}                
\def\aaps{A\&AS}              
\def\mnras{MNRAS}             
               
\ifoldfss
  \ifCUPmtlplainloaded \else
    \NewTextAlphabet{textbfit} {cmbxti10} {}
    \NewTextAlphabet{textbfss} {cmssbx10} {}
    \NewMathAlphabet{mathbfit} {cmbxti10} {} 
    \NewMathAlphabet{mathbfss} {cmssbx10} {} 
  \fi
  \ifAMStwofonts
    \ifCUPmtlplainloaded \else
      \NewSymbolFont{upmath} {eurm10}
      \NewSymbolFont{AMSa} {msam10}
      \NewMathSymbol{\upi}     {0}{upmath}{19}
      \NewMathSymbol{\umu}     {0}{upmath}{16}
      \NewMathSymbol{\upartial}{0}{upmath}{40}
      \NewMathSymbol{\leqslant}{3}{AMSa}{36}
      \NewMathSymbol{\geqslant}{3}{AMSa}{3E}

      \let\leq=\leqslant 
       
    \fi
  \fi
\fi 

\ifnfssone
  \newmathalphabet{\mathit}
  \addtoversion{normal}{\mathit}{cmr}{m}{it}
  \addtoversion{bold}{\mathit}{cmr}{bx}{it}
  \newmathalphabet{\mathbfit} 
  \addtoversion{normal}{\mathbfit}{cmr}{bx}{it}
  \addtoversion{bold}{\mathbfit}{cmr}{bx}{it}
  \newmathalphabet{\mathbfss} 
  \addtoversion{normal}{\mathbfss}{cmss}{bx}{n}
  \addtoversion{bold}{\mathbfss}{cmss}{bx}{n}
  \ifAMStwofonts
    \ifCUPmtlplainloaded \else
      \UseAMStwoboldmath
      \makeatletter
      \new@mathgroup\upmath@group
      \define@mathgroup\mv@normal\upmath@group{eur}{m}{n}
      \define@mathgroup\mv@bold\upmath@group{eur}{b}{n}
      \edef\UPM{\hexnumber\upmath@group}
      \new@mathgroup\amsa@group
      \define@mathgroup\mv@normal\amsa@group{msa}{m}{n}
      \define@mathgroup\mv@bold\amsa@group{msa}{m}{n}
      \edef\AMSa{\hexnumber\amsa@group}
      \makeatother
      \mathchardef\upi="0\UPM19
      \mathchardef\umu="0\UPM16
      \mathchardef\upartial="0\UPM40
      \mathchardef\leqslant="3\AMSa36
      \mathchardef\geqslant="3\AMSa3E

      \let\leq=\leqslant 

    \fi
  \fi
\fi 

\ifnfsstwo
  \DeclareMathAlphabet{\mathbfit}{OT1}{cmr}{bx}{it}
  \SetMathAlphabet\mathbfit{bold}{OT1}{cmr}{bx}{it}
  \DeclareMathAlphabet{\mathbfss}{OT1}{cmss}{bx}{n}
  \SetMathAlphabet\mathbfss{bold}{OT1}{cmss}{bx}{n}
  \ifAMStwofonts
    \ifCUPmtlplainloaded \else
      \DeclareSymbolFont{UPM}{U}{eur}{m}{n}
      \SetSymbolFont{UPM}{bold}{U}{eur}{b}{n}
      \DeclareSymbolFont{AMSa}{U}{msa}{m}{n}
      \DeclareMathSymbol{\upi}{0}{UPM}{"19}
      \DeclareMathSymbol{\umu}{0}{UPM}{"16}
      \DeclareMathSymbol{\upartial}{0}{UPM}{"40}
      \DeclareMathSymbol{\leqslant}{3}{AMSa}{"36}
      \DeclareMathSymbol{\geqslant}{3}{AMSa}{"3E}

      \let\leq=\leqslant 

    \fi
  \fi
\fi 

\ifCUPmtlplainloaded \else
  \ifAMStwofonts \else 
    \def\upi{\pi}
    \def\umu{\mu}
    \def\upartial{\partial}
  \fi
\fi

\title[Extended Gas in Seyfert Galaxies]{Extended Gas in Seyfert 
Galaxies: Near-Infrared Observations of 15 Active Nuclei}
\author[C. Winge et al.]
       {Cl\'audia Winge\,$^{1,}$\thanks{Visiting Astronomer at  the Cerro
Tololo Interamerican Observatory, operated by the Association of
Universities for Research in Astronomy, Inc.  under contract with the
National Science Foundation}\thanks{CNPq Fellowship}, 
        Thaisa Storchi-Bergmann\,$^{1,\bigstar}$, 
        Martin J. Ward\,$^{2,\bigstar}$ 
\newauthor
        and  Andrew S. Wilson\,$^{3,4,\bigstar}$\\
$^1$\,Instituto de F\'{\i}sica, UFRGS, Av. Bento Gon\c{c}alves, 9500, 
C.P. 15051. CEP\,91501-970, Porto Alegre, RS, Brazil.\\
$^2$\,Dept. of Physics and Astronomy, University of Leicester, 
University Road, Leicester LE1 7RH, England\\
$^3$\,Astronomy Program, University of Maryland, College Park, 
MD\,20742, USA\\
$^4$\,Space Telescope Science Institute, 3700 San Martin Dr., 
Baltimore, MD\,21218, USA}
\date{Accepted December 8, 1999; Received August 24, 1999}

\pagerange{\pageref{firstpage}--\pageref{lastpage}}
\pubyear{2000}

\begin{document}

\maketitle

\label{firstpage}

\begin{abstract}
Results from an analysis of low resolution (R\app 250) near-IR
long-slit spectra covering simultaneously the I, J, H, and K bands, for
a sample of 15 Seyfert galaxies and the NGC\,5253 starburst nucleus,
are presented. The Seyfert galaxies were selected as presenting
`linear' or cone-like high excitation emission line in the optical,
most probably due to the collimation of the central source's radiation
by a dusty molecular torus. Our goal was to look for signatures of this
torus, and to investigate the gaseous distribution, excitation and
reddening.  The strongest emission lines detected are usually
\bhen\ and \bsiiia, followed by \pabeta. In some cases, \bfeiia\ and
1.64\mic\ are also seen.  \bfeiia\ and \hhl\ are detected in some of
the higher resolution spectra obtained for five galaxies.  The emission
lines are spatially extended in most cases, and we have used the
[\feii]/\pabeta\ ratio as a measure of the gaseous excitation in
Mrk\,573, NGC\,1386, and NGC\,7582. Values for this ratio between 1.5
and 6 are found, suggesting excitation of [\feii] by X-rays or shock
waves in some regions. Broad permitted lines are observed in three
Seyfert 1 galaxies. Nuclear \pabeta\ in NGC\,1365, and possibly nuclear
\brgama\ in Mrk\,573, are also broad.

{}From analysis of the spatial distribution of the continuum \jmh\ and
\hmk\ colours derived from our spectra, we find redder colours for the
nucleus than the nearby bulge in most of the Seyfert 2s observed.
Comparison with models including emission from dust and stars shows
that hot (T \app\ 1000 K) dust emission dominates the nuclear continuum
in NGC\,1365, NGC\,2110, NGC\,3281, NGC\,7582, and ESO362-G18. In
NGC\,1386, NGC\,5643, and NGC\,5728 the main contributor is the
underlying stellar population, combined with some foreground reddening
and/or cooler dust emission. In a few cases, the \jmh\ colours on
opposite sides of the nucleus differ by 0.3 -- 0.8 mag, an effect that
we interpret as partly due to differences in the local stellar population, 
and possibly extinction gradients.

\end{abstract}


\section{Introduction} \label{sec_intro}

In the unified model for active galactic nuclei (AGN) the nuclear
engine is surrounded by an optically thick, dusty toroidal structure,
which collimates the escaping photons and, at the same time, hides the
true nucleus from our line of sight (Antonucci 1993).  Observational
evidence for such a model come from optical spectropolarimetry, which
reveals that a number of Seyferts 2s show broad emission lines and blue
continuum in polarized light (Miller \& Goodrich 1990; Tran 1995), and
emission-line imaging, which shows elongated morphologies for the
extended emission line regions, with striking conical or bi-conical
structures that can be traced down to a few tens of parsecs from the
nucleus and which are generally aligned with the radio ejecta (Pogge
1989; Storchi-Bergmann \& Bonatto 1991; Wilson \& Tsvetanov 1994).

There are few observational constraints on the geometry of the torus,
although red features and dust lanes (e.g., Wilson et al. 1993; Simpson
et al. 1996b) aligned perpendicular to the radio/extended line emission
axis are seen in some high resolution images. Hot dust (T  \app\ 1200
K) has been found to be an important contributor to the near-infrared
nuclear continuum of Seyfert galaxies (Glass \& Moorwood 1985;
Kotilainen et al. 1992; Alonso-Herrero et al. 1998). Theoretical models
(Krolik \& Begelman 1988) show that for the dust to reach this
temperature, the inner edges of the torus must be only a few parsecs
from the nuclear source. The total extent of the torus, however, is not
well constrained, and may reach several tens or even a hundred
parsecs.  Warm dust (T \dapp\ 600 -- 800 K) is presumably responsible
for the observed mid-infrared flux (Pier \& Krolik 1993).

Infrared spectroscopy of a number of Seyfert 2 galaxies has revealed
broad components in the hydrogen emission-lines (e.g., Veilleux,
Goodrich \& Hill 1997 and references therein), thus confirming, at
least in those particular cases, the presence of an obscured broad-line
region. The torus is expected to contain warm molecular hydrogen, whose
vibration-rotational transitions, especially \hh\ $\nu$=1$-$0 S(1),
could be used as a tracer of the size and geometry of the toroidal
structure (Blietz et al. 1994; Moorwood et al. 1996; Marco, Alloin \&
Beuzit 1997). On the other hand, while the \hh\ emission is expected to
be extended perpendicular to the collimation/radio emission axis,
emission lines  from ionized gas (such as \bfeiia) should be emitted
preferentially along the collimation axis, either due to
photoionization by the hard photons from the central source or by
shocks from the interaction of the radio jet with the ambient gas.

To address these issues, we obtained long-slit spectra in the
near-infrared I, J, H, and K bands of a sample of Seyfert galaxies with
known elongated or bi-conical high-excitation optical emission, which,
in the unified model, is a direct result of the presence of the
collimating torus. We also observed NGC\,5253, a classical \ion{H}{ii}
galaxy. Our first results, on NGC\,2110 and the Circinus galaxy, have
already been published (Storchi-Bergmann et al.  1999, hereafter Paper
I), and in this paper we present the data for the remaining galaxies
observed.

\section{Data Reduction} \label{sec_obs}

Long-slit spectra of the sample galaxies were obtained using the
Infrared Spectrograph (IRS) on the 4m `Blanco' telescope of the Cerro
Tololo Interamerican Observatory in October-November, 1995 and
February-March, 1996. The detector used was a 256\vzs 256 InSb device,
with a 0\farcs363 per pixel scale. The useful slit length was about
15\arcsec, and the slit width was either 1\arcsec\ or 1\farcs7,
depending on the seeing, which was generally between these two values.
The majority of the spectra were obtained using a low-resolution (R\app
250), cross-dispersed grating (hereafter, the XD grating), which splits
the complete 0.9 -- 2.3 \mic\ spectra into four or five segments,
roughly coincident with the R, I, J, H, and K spectral bands. A few
galaxies were also observed using either a 75 lines per mm grating,
with resolution R\app 700 (4 pixels), hereafter the LR grating, or a
210 lines per mm grating, R\app 2000, hereafter the HR grating.

The log of observations is listed in Table~\ref{tab_log}. We observed
most of the galaxies with the slit aligned along the position angle
(PA) of the major axis of the inner isophotes in narrow-band
([\ion{O}{iii}]) images or along the radio axis (identified by an `r'
in Column 3). In a few cases, spectra were also obtained along the
perpendicular direction (`p') or,  for the Seyfert 1 objects, along
arbitrary PAs.  For NGC\,5253 the slit was positioned along the major
axis of the \halfa\ emission structure in the inner \app\ 25\arcsec.
Futher details of the individual spectra are given in
Section~\ref{sec_res}.  To obtain the linear spatial scales in
Table~\ref{tab_log}, we adopted \hbc\ = 75 \kms\,Mpc$^{-1}$.

\begin{table*}
\begin{minipage}{150mm}
\caption{Log of Observations \label{tab_log}}
\begin{tabular}{lcccccc}
\hline\hline
Object       & Date        &  P.A.   & Grating/Band &  Exp. Time & Slit 
width        & Spatial scale \\ 
             &             &(\degr)&              &  (sec)     & 
(\arcsec)    & (pc/\arcsec)  \\
\hline
NGC\,526A    & 1995 Oct 30 & 123     & XD    & 1100 & 1.1 & 373 \\
NGC\,1097    & 1995 Oct 31 & 77      & LR~J  & 1200 & 1.1 & 82  \\
NGC\,1365    & 1995 Oct 30 & 130     & XD    & 600  & 1.1 & 106 \\
             & 1995 Oct 31 & 145     & LR~J  & 2400 & 1.1 &     \\
NGC\,1386    & 1996 Feb 28 & 0       & XD    & 800  & 1.1 & 56  \\
NGC\,2110    & 1996 Feb 28 & 350 (r) & XD    & 800  & 1.1 & 148 \\
NGC\,3281     & 1996 Feb 28 & 45     & XD    & 800  & 1.1 & 224 \\
NGC\,4388\,$^a$  & 1996 Feb 27 & 13  & XD    & 1200 & 1.1 & 78 \\
             & 1996 Feb 27 & 90  (p) & XD    & 800  & 1.1 &     \\
             & 1996 Feb 29 & 13      & HR~J  & 2200 & 1.7 &     \\
             & 1996 Feb 29 & 90  (p) & HR~J  & 2200 & 1.7 &     \\
NGC\,5253    & 1996 Feb 28 & 43      & XD    & 400  & 1.1 & 26  \\
NGC\,5643    & 1996 Feb 28 & 90      & XD    & 800  & 1.1 & 78  \\
NGC\,5728    & 1996 Feb 28 & 20  (p) & XD    & 400  & 1.1 & 180 \\
             & 1996 Feb 28 & 110     & XD    & 800  & 1.1 &     \\
NGC\,7582    & 1995 Oct 30 & 203     & XD    & 900  & 1.1 & 102 \\
IC\,5063\,$^b$ & 1995 Nov 01 & 90    & LR~K  & 2520 & 1.1 & 220 \\
ESO\,362-G18 & 1995 Oct 30 & 68  (p) & XD    & 900  & 1.1 & 245 \\
Fairall\,9   & 1995 Oct 30 & 123     & XD    & 600  & 1.1 & 911 \\
Mrk\,509     & 1995 Oct 30 & 0       & XD    & 600  & 1.1 & 667 \\
Mrk\,573     & 1995 Oct 31 & 125 (r) & LR~J  & 3000 & 1.1 & 335 \\
             & 1995 Nov 01 & 35  (p) & LR~K  & 3240 & 1.1 &     \\
\hline
\end{tabular}

\smallskip
$^a$\ Bad seeing (\app\ 2\farcs5) for the HR spectra at PA=13\degr.\\
$^b$\ Clouds.
\end{minipage}
\end{table*}

Both XD and LR/HR data were reduced using {\sc IRAF}\,\footnote{{\sc
IRAF} is distributed by the National Optical Astronomy Observatories,
which are operated by the Association of Universities for Research in
Astronomy, Inc., under cooperative agreement with the National Science
Foundation.} scripts kindly provided by R. Elston at CTIO (available at
the CTIO ftp archive), and followed standard procedures.  The reduction
of the LR and HR data has already been described in Paper I.  The only
difference here is that, after bias subtraction, flatfielding, and sky
subtraction, removal  of the atmospheric absorption features  and flux
calibration were performed on the full frame, instead of on the
extracted spectra.

For the XD data, the fundamental difference from the standard IR data
reduction is the tracing of the distortion across the slit (the
wavelength direction).  This was done using the spectrum of a bright
star or pinhole, which was traced in each `band' with a 4th order
Legendre polynomial in the {\tt aptrace} task. Similarly, the
flat-fields were created using {\tt apnormalize} rather than {\tt
response}, and then the individual segments of each image
(corresponding approximately to the five colour bands) were
flat-fielded, and `straightened' using the solution previously found
with the tracing star/pinhole spectrum. Wavelength calibration was
performed using an HeAr lamp, with different solutions for each
`band'.  Low (2 -- 4) order Legendre polynomials were fitted to a
sample of 4 (K) to 12 (R) lines, resulting in the spectral intervals,
pixel scales, and rms residuals of the fits listed in
Table~\ref{tab_cdwav}.  Atmospheric absorption correction and flux
calibration followed the same procedures as for the LR/HR data.

\begin{table}
\caption{Cross-dispersed Grating Wavelength Calibrations \label{tab_cdwav}}
\begin{tabular}{lcccc}
\hline\hline
Run & Band & Range  & Pixel & rms   \\
    &      & (\mic) & (\AA) & (\AA) \\
\hline
1995 Oct 30 & I & 0.802 -- 1.195 & 16 & 2  \\
            & J & 0.961 -- 1.435 & 19 & 7  \\
            & H & 1.199 -- 1.792 & 23 & 10 \\
            & K & 1.601 -- 2.364 & 30 & 32 \\
1996 Feb 27 & I & 0.865 -- 1.260 & 16 & 3  \\
            & J & 1.038 -- 1.513 & 19 & 3  \\
            & H & 1.296 -- 1.890 & 24 & 7  \\
            & K & 1.602 -- 2.691 & 43 & 66 \\
1996 Feb 28 & R & 0.772 -- 0.999 & 13 & 26 \\
            & I & 0.851 -- 1.226 & 16 & 2  \\
            & J & 0.972 -- 1.447 & 19 & 5  \\
            & H & 1.213 -- 1.806 & 24 & 4  \\
            & K & 1.616 -- 2.404 & 31 & 16 \\
\hline
\end{tabular}
\end{table}

\section{Results} \label{sec_res}

The integrated spectra of the sample galaxies, corresponding to the
co-added emission within a 7 -- 8\arcsec\ spatial window are shown in
Figures \ref{fig_specxd}a to \ref{fig_specxd}e for the XD data. For the
LR/HR data, the extraction windows were as listed on
Table~\ref{tab_hrflx} and the resulting integrated spectra are shown in
Figures~\ref{fig_spechr}a and \ref{fig_spechr}b. Inside the atmospheric
absorption bands, the XD data was masked out when the signal was below
10 percent of the peak transmission in the uncalibrated frames. The
main emission lines are indicated, and total line fluxes are listed in
Tables~\ref{tab_xdflx} and \ref{tab_hrflx}.  In the XD sub-sample, the
strongest lines are the \bhepg\ blend and \bsiiia\ emission.  The
uncertainties in the integrated fluxes are dominated by the continuum
placement, and we estimate them to be of the order 20 -- 25\% for the
XD spectra, and 10 -- 20\% for the LR/HR data.

Table \ref{tab_hrfw} lists the full width at half maximum (FWHM),
corrected for instrumental resolution, of the lines observed at
different positions for the four objects with HR/LR spectra. The errors
are due to the uncertainty in the placement of the continuum and
represent maximum values.  Due to its higher S/N ratio, the emission
lines in a spectrum of the planetary nebula NGC\,7009 were used as the
reference (instrumental) profile for the LR grating. These values were
found to be in very good agreement with the FWHM of the sky lines. For
the HR grating the instrumental profile was given by the HeAr
calibration lamp lines.

\begin{figure}
\psfig{file=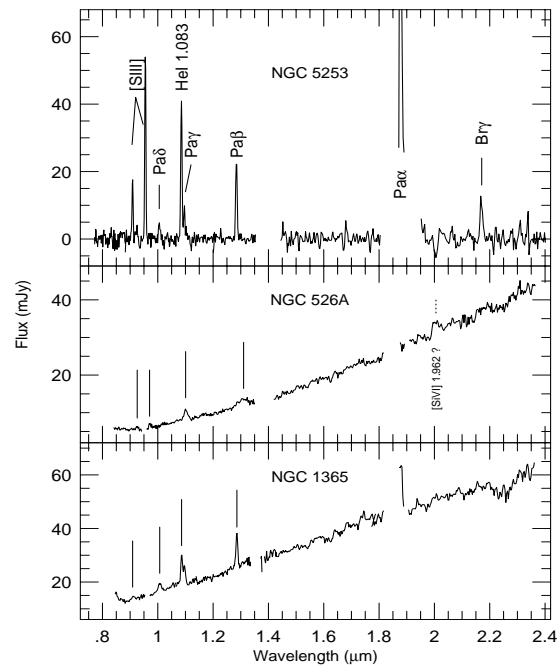,height=10.5cm,width=0.5\textwidth}
\caption{(a) Integrated spectra of the galaxies observed with 
the XD grating. NGC\,5253, NGC\,526A, and NGC\,1365}
\label{fig_specxd}
\end{figure}

\begin{figure}
\psfig{file=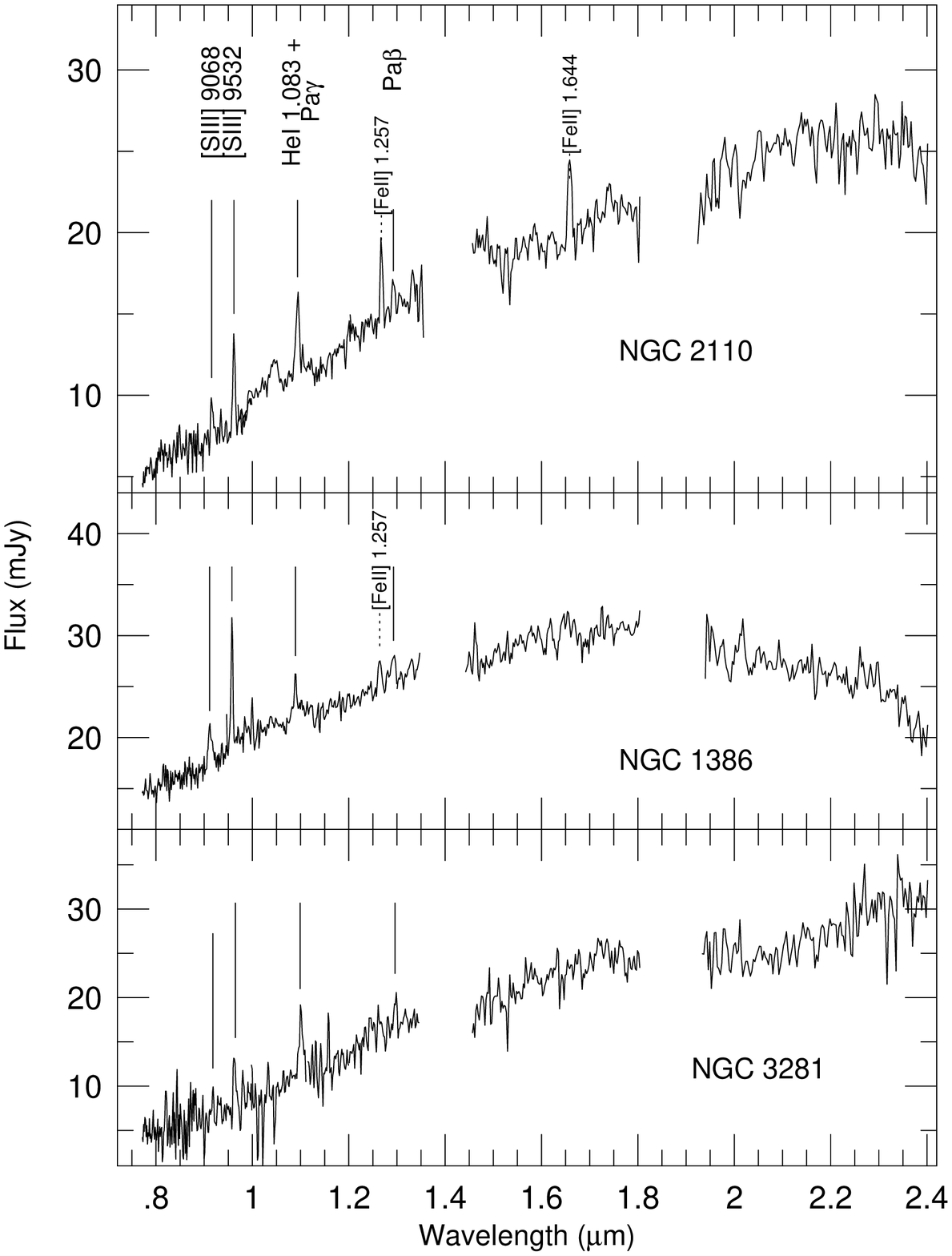,height=10.5cm,width=0.5\textwidth}
\contcaption{\ (b) NGC\,2110, NGC\,1386, and NGC\,3281}
\end{figure}

\begin{figure}
\psfig{file=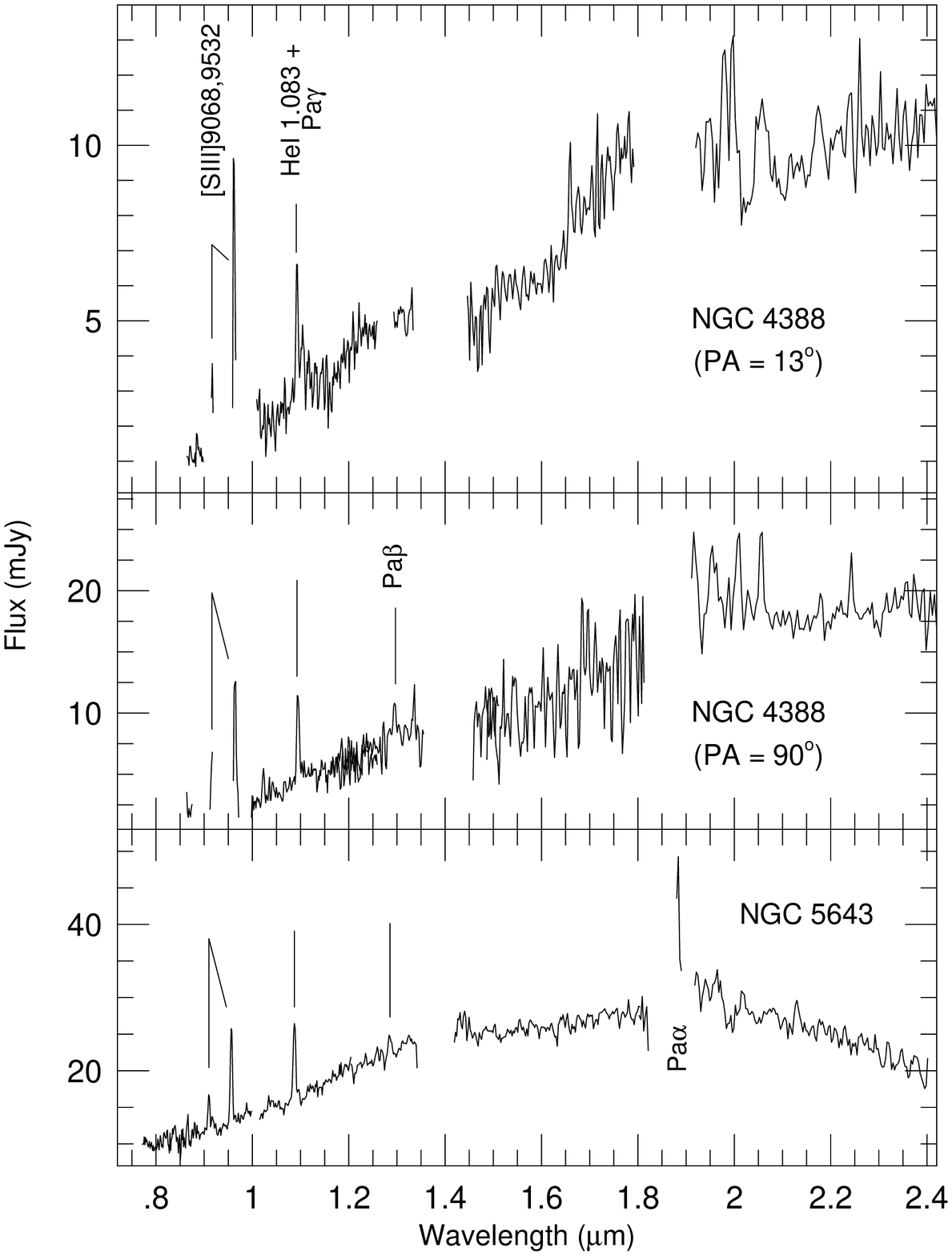,height=10.5cm,width=0.5\textwidth}
\contcaption{\ (c) NGC\,4388 (PA=13\degr\ and 90\degr), and NGC\,5643}
\end{figure}

\begin{figure}
\psfig{file=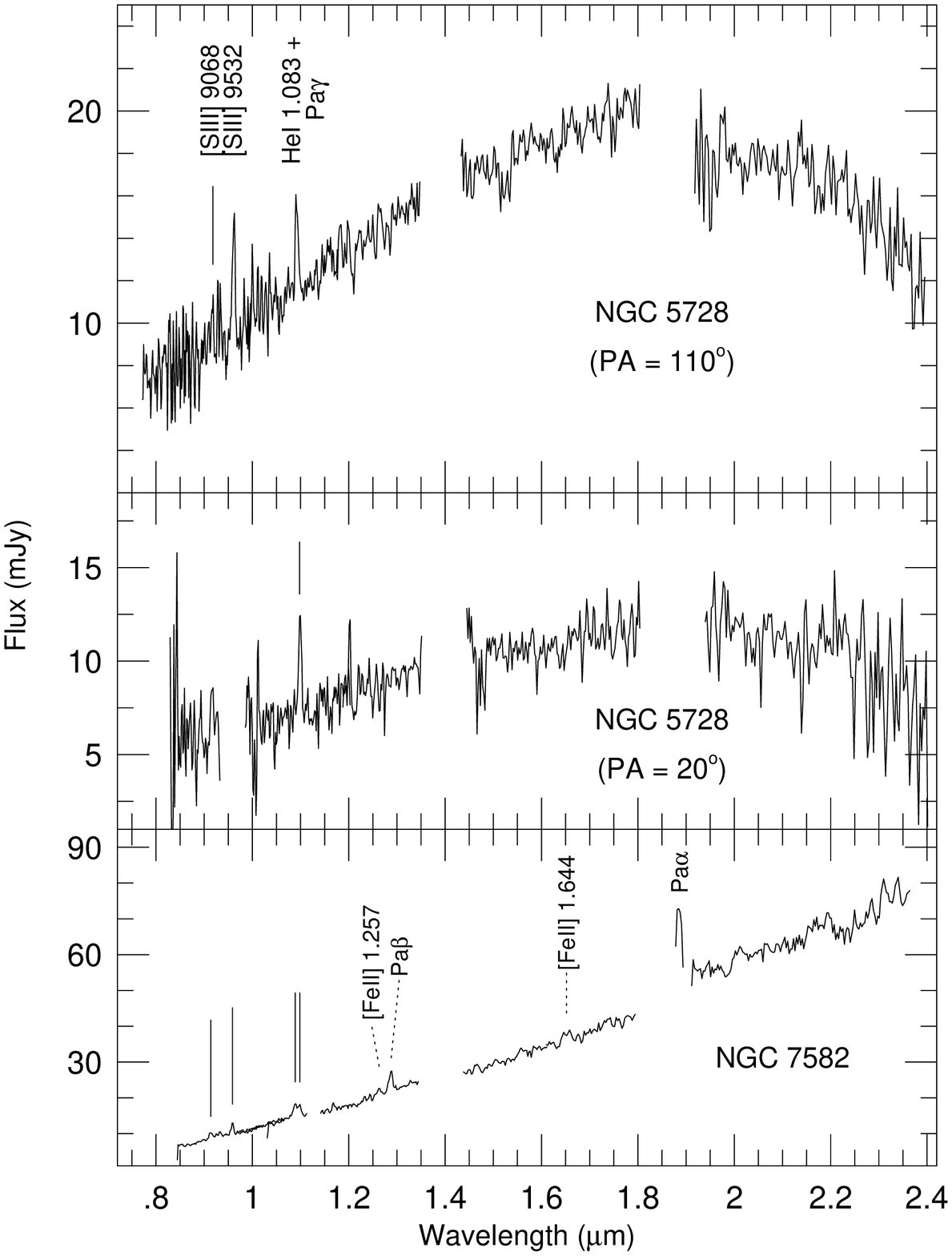,height=10.5cm,width=0.5\textwidth}
\contcaption{\ (d) NGC\,5728 (PA=110\degr\ and 20\degr), and NGC\,7582}
\end{figure}

\begin{figure}
\psfig{file=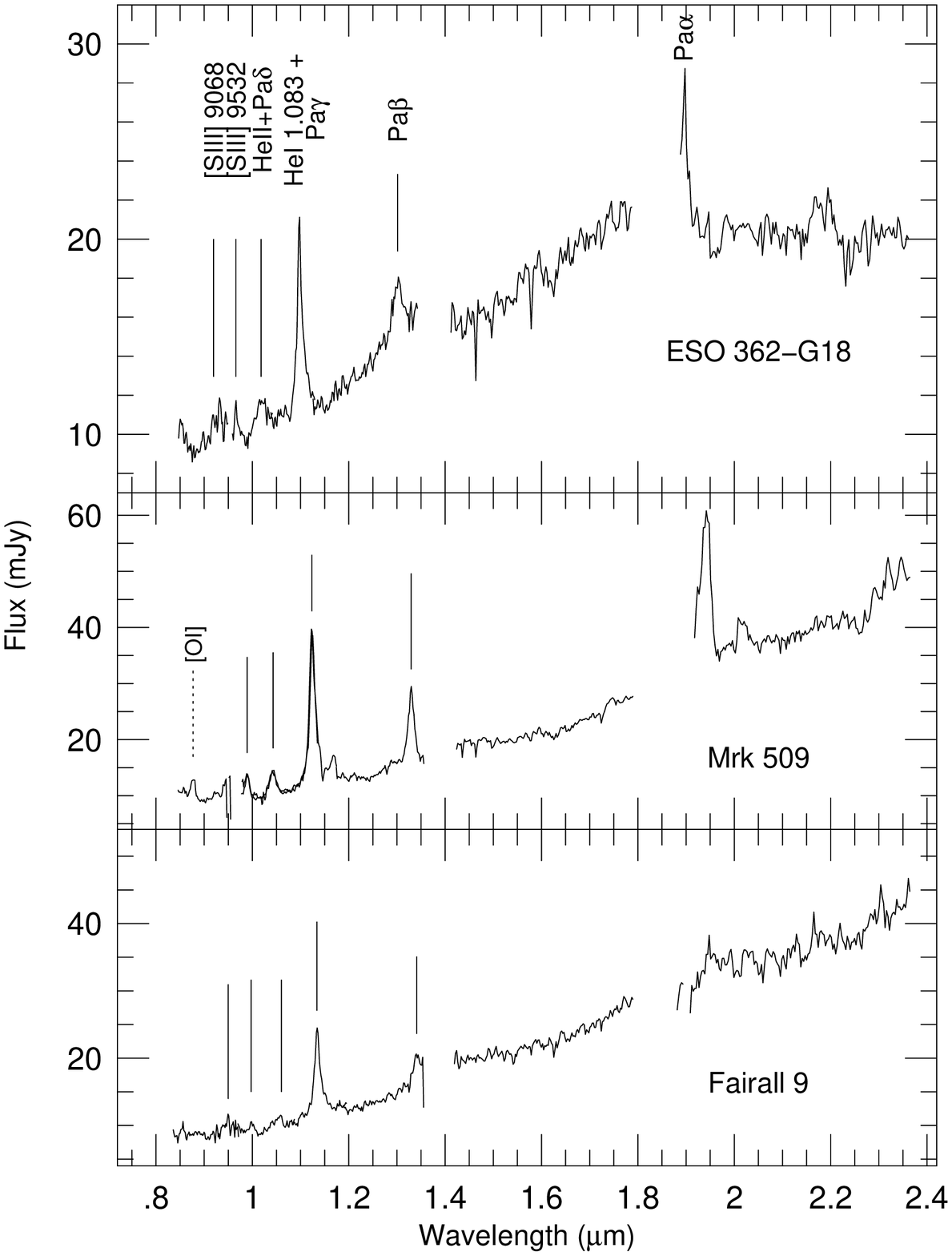,height=10.5cm,width=0.5\textwidth}
\contcaption{\ (e) ESO\,362-G18, Mrk\,509, and Fairall\,9}
\end{figure}

\begin{table}
\caption{Integrated line fluxes from the spectra obtained with 
the XD grating  (in units of \tento{-14} \ergcs). 
\label{tab_xdflx}}
\begin{tabular}{lccc}
\hline\hline
Object & Line ID & Flux \\
\hline
NGC\,526A & [\ion{S}{iii}]\,9532   &  3.9  \\
          & \bhepg                 & 11.7  \\
          & \pabeta                & 11.5  \\
NGC\,1365 & [\ion{S}{iii}]\,9532   & 11.5  \\
          & \padelta               &  3.1  \\
          & \ion{He}{i}\,1.083     & 26.6  \\
          & \pagama                & 14.7  \\
          & \pabeta                & 20.2  \\
NGC\,1386 & [\ion{S}{iii}]\,9068   &  6.5  \\
          & [\ion{S}{iii}]\,9532   & 16.1  \\
          & \bhepg                 &  6.6  \\
          & [\feii]\,1.26          &  5.3  \\
          & \pabeta                &  3.5  \\
NGC\,2110 &  [\ion{S}{iii}]\,9068  &  4.9  \\
          &  [\ion{S}{iii}]\,9532  & 11.8  \\
          & \bhepg                 & 10.1  \\
          & [\feii]\,1.26          &  7.0  \\
          & \pabeta                &  3.4  \\
          & [\feii]\,1.64          &  6.2  \\
NGC\,3281 &  [\ion{S}{iii}]\,9068  &  7.6  \\
          &  [\ion{S}{iii}]\,9532  & 24.1  \\
          & \bhepg                 &  7.9  \\
NGC\,4388 &  [\ion{S}{iii}]\,9532  &  8.4  \\
 (PA=13)  & \bhepg                 &  5.7  \\
NGC\,4388 &  [\ion{S}{iii}]\,9532  & 23.1  \\
 (PA=90)  & \bhepg                 & 11.8  \\
          & \pabeta                &  5.6  \\
NGC\,5253 & [\ion{S}{iii}]\,9068   & 33.6  \\
          & [\ion{S}{iii}]\,9532   & 94.3  \\
          & \padelta               &  8.7  \\
          & \ion{He}{i}\,1.083     & 64.8  \\
          & \pagama                & 13.2  \\
          & \pabeta                & 31.1  \\
          & \palfa                 & 121.7 \\
          & \brgama                &  9.1  \\
NGC\,5643 & [\ion{S}{iii}]\,9068   &  8.7  \\
          & [\ion{S}{iii}]\,9532   & 23.3  \\
          & \bhepg                 & 17.3  \\
          & \pabeta                &  3.5  \\
NGC\,5728 &  [\ion{S}{iii}]\,9532  & 14.4  \\
(PA=110)  & \bhepg                 &  9.2  \\
NGC\,5728 &   [\ion{S}{iii}]\,9532 & 11.9  \\
(PA=20)   &   \bhepg               & 10.2 \\
NGC\,7582 &  [\ion{S}{iii}]\,9532  &  5.7  \\
          & \bhepg                 & 14.7  \\
          & [\feii]\,1.26          &  2.1  \\
          & \pabeta                &  7.8  \\
          & [\feii]\,1.64          &  5.0  \\
ESO\,362-G18 & \bhepg              & 35.0  \\
             & \pabeta             & 10.1  \\
             & \palfa              & $>$10.7 \\
Fairall\,9 &  [\ion{S}{iii}]\,9532 &  9.0    \\
           & \bhepd                &  7.7    \\
           & \bhepg                & 51.5    \\
           & \pabeta               & 15.3    \\
           & \palfa                & $>$18.1 \\
Mrk\,509  & [\ion{S}{iii}]\,9532   & 13.3    \\
          & \bhepd                 & 21.9    \\
          & \bhepg                 & 108.0   \\
          & \pabeta                & 33.0    \\
          & \palfa                 & $>$59.1 \\
\hline
\end{tabular}
\end{table}

\begin{figure}
\psfig{file=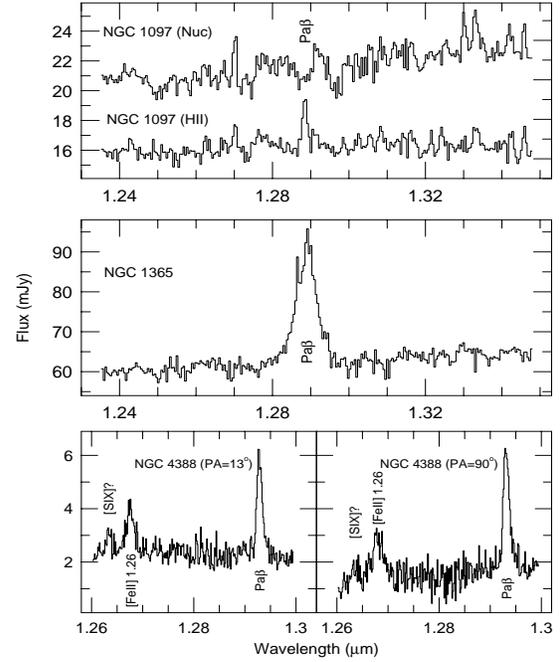,height=10.5cm,width=0.5\textwidth}
\caption{Integrated spectra of the sample galaxies observed with the
LR/HR gratings (a) NGC\,1097 -- LR/J, NGC\,1365 -- LR/J, and NGC\,4388
-- HR/J (PA = 13\degr\ and 90\degr). The right hand scale on the top
panel is the flux scale for the HII region spectrum in NGC\,1097.}
\label{fig_spechr}
\end{figure}

\begin{figure}
\psfig{file=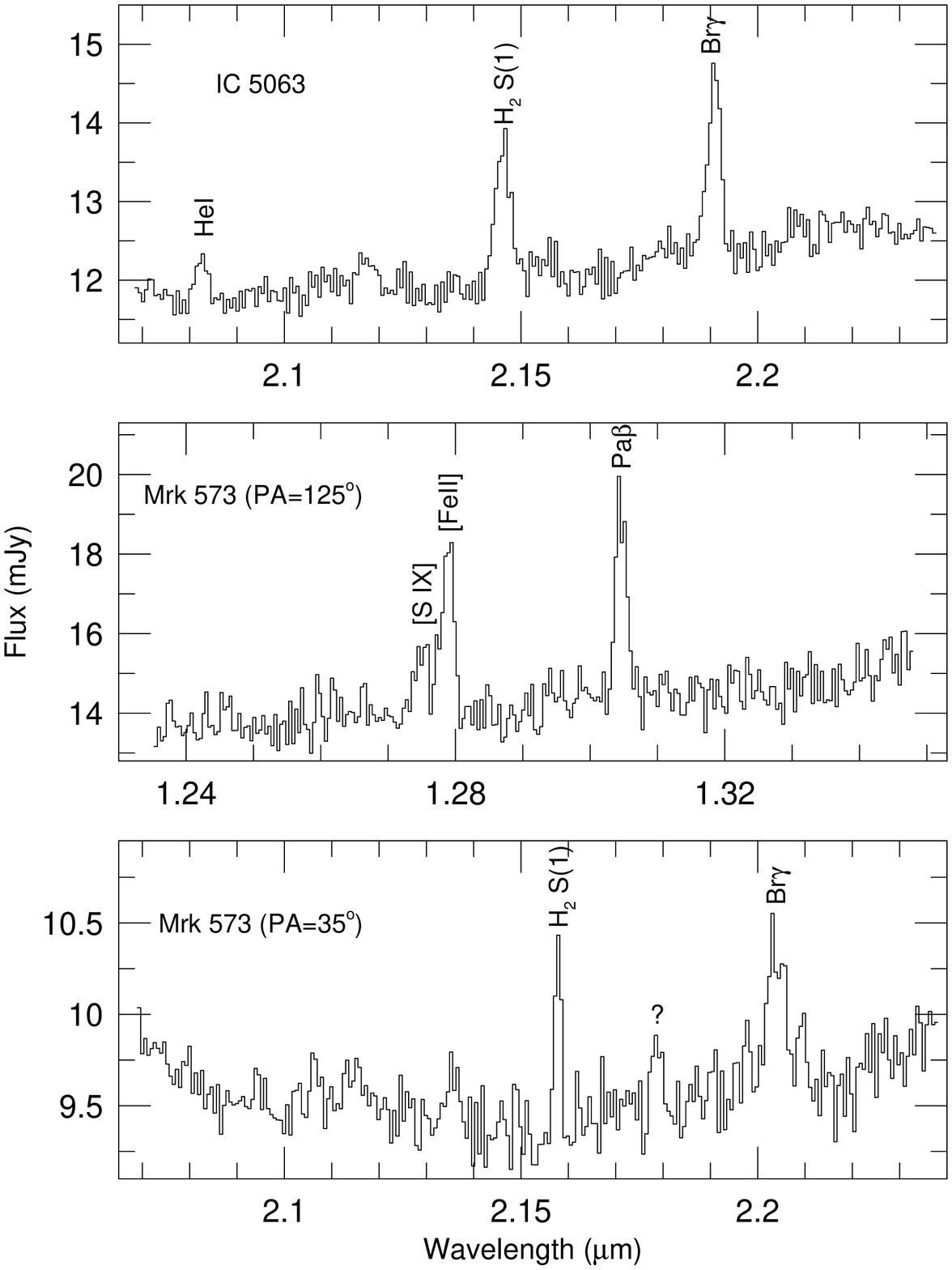,height=10.5cm,width=0.5\textwidth}
\contcaption{\ (b) IC\,5063 -- LR/K, Mrk\,573 -- LR/J 
(PA = 125\degr), and Mrk\,573 -- LR/K (PA = 35\degr)}
\end{figure}

\begin{table}
\caption{Integrated line fluxes from the spectra taken with 
the LR/HR gratings (in units of \tento{-14} \ergcs) 
\label{tab_hrflx}}
\begin{tabular}{lccc}
\hline\hline
Object & Window (\arcsec) & Line ID & Flux  \\
\hline
NGC\,1365 &  10.96 & \pabeta (broad) & 20.8 \\
          &        & \pabeta (total) & 41.3 \\
NGC\,4388 &   3.65 & [\ion{S}{ix}]   & 2.2  \\
(PA=13)   &        &  [\feii]\,1.26  & 6.7  \\
          &        & \pabeta         & 10.0 \\
NGC\,4388 &   4.38 & [\feii]\,1.26   & 8.9  \\
(PA=90)   &        &  \pabeta        & 12.2 \\
IC\,5063  &   5.12 &  \hh\,S(1)      & 0.55 \\
          &        &   \brgama       & 0.45 \\
Mrk\,573  &   8.05 & [\ion{S}{ix}]   & 0.89 \\
(PA=125)  &        & [\feii]\,1.26   & 1.89 \\
          &        & \pabeta         & 2.03 \\
Mrk\,573  &   5.12 & \hh\,S(1)       & 0.14 \\
(PA=35)   &        & \brgama         & 0.37 \\
\hline
\end{tabular}

\smallskip
The two entries for \pabeta\ in NGC\,1365 correspond to the
fluxes in the broad component only (FWHM \dapp\ 1500 \kms; see Table~\ref{tab_hrfw}) and in all the
emission line, respectively.
\end{table}

\subsection{Extended Emission and Line ratios} \label{sec_profiles}

Due to the low spectral resolution of the XD spectra, we opted to
extract the emission line spatial distributions directly from the
two-dimensional frames, rather than extracting a number of spectra
along the slit and measuring the individual line fluxes as done in
Paper I.  The spatial profiles were obtained by extracting and
co-adding all columns inside a spectral window wide enough to contain
all the line emission, and the underlying continuum subtracted by
linear interpolation between two windows at each side of the emission
line. The errors were estimated by creating an image with the same ratio
of the rms to the total raw counts ($\sqrt{N}/N$) in each pixel on the
uncalibrated frames, then processing those images with the same procedure
as used for the actual data. 

With the exception of NGC\,5253, where the geometric centre of the slit
was arbitrarily set as the origin of the spatial axis, the centroid of
a nearby continuum spatial profile was used as the origin of the
emission line spatial distributions. No correction for foreground
(Galactic) reddening was applied.

In most XD spectra, the emission lines with enough S/N ratio to allow
the construction of meaningful spatial profiles were \bsiiia, the
\bhepg\ blend and \pabeta. The available emission lines allowed the
determination of the internal reddening only in the \ion{H}{ii}\ galaxy
NGC\,5253, where we found average reddening corrected values of 3.1 and
4.7 for (\bhepg)/\pabeta, and \bsiiia/\pabeta\ ratios, respectively.
For the other galaxies, a good determination of the narrow emission
line reddening could not be obtained, and any comparison among the
resulting line ratios should be considered with caution. Some
regularity was observed between the Seyfert 1 galaxies, which presented
integrated line ratios (\bhepg)/\pabeta\ \dapp\ 3.4, and
\bsiiia/\pabeta\ \dapp\ 0.4. For the Seyfert 2 nuclei, these same
ratios span the ranges 1.0 $\lesssim$\ (\bhepg)/\pabeta\ $\lesssim$\ 4.9
and 0.6 $\lesssim$\ \bsiiia/\pabeta\ $\lesssim$\ 6.7.

We now describe the spatial distribution of the emission lines for each
galaxy, comparing the light profiles in the emission lines with that of
a reference star,  and also the spatial variations in the available
line ratios. For reasons of clarity in the figures, the emission line spatial profiles shown below have been rebbined to 0\farcs74 per pixel.

\begin{table}
\caption{Intrinsic line widths (\kms). The upper block are measurements 
with the LR grating, the lower with the HR grating. 
\label{tab_hrfw}}
\begin{tabular}{llcl}
\hline\hline
Object & Position(\arcsec) & Line & FWHM \\
\hline
NGC\,1365  & Nuc             & \pabeta\,(N)  & $<$ 450\\
           &                 & \pabeta\,(B)  & 1397 $^{+30}_{-15}$\\
           &                 & \pabeta\,(VB) & 2868 $^{+349}_{-729}$\\
           & 4.02 NW         & \pabeta       & 159 $^{+48}_{-52}$\\
           & 4.02 SE         & \pabeta       & 210 $^{+53}_{-69}$\\
IC\,5063   & Nuc             & \brgama       & $<$ 400\\
           &                 & \hh\,S(1)     & $<$ 400\\
           & 1.46 NW         & \brgama       & $<$ 400 \\
           &                 & \hh\,(S1)     & 483 $^{+82}_{-73}$\\
Mrk\,573 J & Nuc             & [\ion{S}{ix}] & $<$ 450 \\
           &                 & \bfeiia       & 323 $^{+54}_{-46}$\\
           &                 & \pabeta       & $<$ 450\\
Mrk\,573 J & 1.46 NW         & \bfeiia       & $<$ 450\\
           &                 & \pabeta       & 182 $^{+127}_{-102}$ \\
           & 1.46 SE         & \bfeiia       & $<$ 450 \\
           &                 & \pabeta       & 323 $^{+43}_{-127}$ \\
           & 3.29 NW         & \pabeta       & $<$ 450\\
           & 3.29 SE         & \pabeta       & $<$ 450\\
Mrk\,573 K & Nuc             & \brgama       & 589 $^{+256}_{-135}$ \\
           &                 & \hh\,S(1)     & $<$ 400\\
NGC\,7009  & (instrumental, LR) &  \pabeta   & 455 $^{+8}_{-13}$\\
           &                    & \brgama    & 403 $^{+10}_{-8}$\\
\hline
NGC\,4388 & Nuc    & \bfeiia     & 297 $^{+32}_{-30}$\\
\ (PA=13) &        & \pabeta     & 299 $^{+69}_{-24}$\\
NGC\,4388 & Nuc    & \pabeta     & 378 $^{+62}_{-40}$\\
\ (PA=90) & 1.46 W & \pabeta     & 171 $^{+48}_{-25}$\\
HeAr      & (instrumental, HR) & & 148 \mm\ 1\\
\hline
\end{tabular}
\end{table}

\begin{figure}
\psfig{file=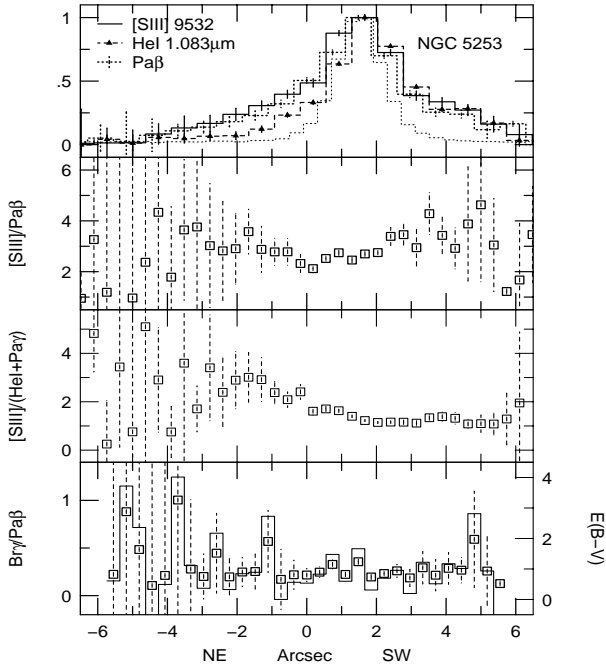,height=10.5cm,width=0.5\textwidth}
\caption{Normalized spatial distributions and line ratios of the main
emission lines from the XD spectrum of NGC\,5253. The dotted line in
the upper panel is the spatial profile of the atmospheric standard star
used in the data reduction and approximately represents the seeing
profile. The full line in the bottom panel represents the value of
\ebv\ (right axis scale) calculated from the \brgama/\pabeta\ ratio 
(squares, left axis scale).}
\label{fig_n5253}
\end{figure}

{\it NGC\,5253:} The spectrum of this \ion{H}{ii}\ galaxy
(Figure~\ref{fig_specxd}a, top panel) was obtained along the PA of the
major axis of the \halfa\ emission structure seen in Fig. 1a of
Calzetti et al. (1997), with essentially no stellar continuum
detected.  The strong emission lines (Fig.~\ref{fig_n5253}) extend
over \app\ 10\arcsec\ and peak at the central clump, corresponding to
the starburst nucleus. To investigate their variation with distance
from the nucleus, we constructed several line ratios. The
\bsiiia/(\bhepg) ratio varies from \app\ 1.2 at 3\arcsec\ SW to
\app\ 2.8 at 3\arcsec\ NE, while the \bsiiia/\pabeta\ is almost flat in
the above interval, 2.78\mm 0.96, rising slightly to the SW. The
average reddening in the region $-$3\arcsec\ $< r <$ 4\farcs5,
calculated from the \brgama/\pabeta\ ratio, assuming a case B intrinsic
value of 0.171 (Osterbrock 1989), is \ebv\ \app\ 0.91\mm 0.15. However,
large variations (as much as \app\ 3 mag in \ebv) are found within a
few pixels, in agreement with the work of Calzetti et al (1997), which
detected a much higher reddening inside the emission clumps than
indicated by the emission line ratios in the integrated optical
spectrum. The (\bhepg)/\pabeta\ ratio, which can be interpreted as a
tracer of the ionization structure, varies from \app\ 2 in the extended
regions to up to 5.4 in the starburst nucleus. Assuming a case B
recombination, and \av\ = 3 -- 6 mag, the contribution of \pagama\ to
the blend varies between 0.45 to 0.36 the \pabeta\ flux. For the above range
of reddening, the observed line ratios translate to 2.6
$\lesssim$\ \bhen/\pabeta\ $\lesssim$\ 3.5, 1.9
$\lesssim$\ \bsiiia/\bhen\ $\lesssim$\ 2.2 (both ratios corrected for the
\pagama\ contribution), 4.7 $\lesssim$\ \bsiiia/\pabeta\ $\lesssim$\ 7.8.
The upper limits are consistent with  \ion{H}{ii}\ regions models (Lumsden \& Puxley 1996).

\begin{figure}
\psfig{file=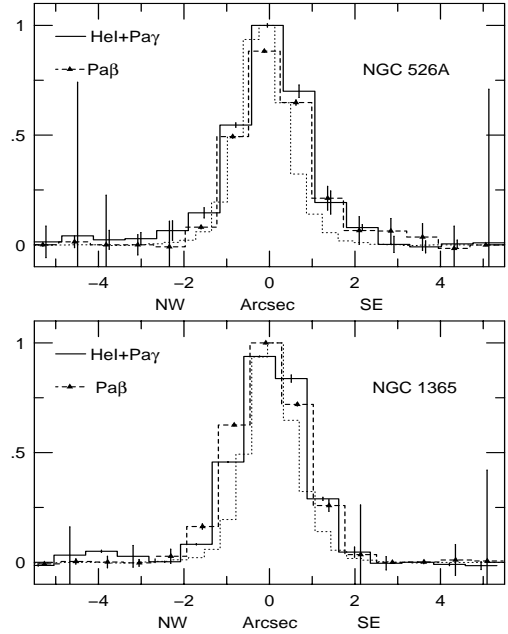,height=10cm,width=0.5\textwidth}
\caption{Spatial profiles of the main emission lines from the XD
spectrum of NGC\,526A (top) and NGC\,1365 (bottom). As for NGC\,5253,
the dotted line is the stellar profile.}
\label{fig_n5_n13}
\end{figure}

{\it NGC 526A:} The XD spectrum (Fig.~\ref{fig_specxd}a, middle) was
obtained along the PA of the major axis of the [\ion{O}{iii}] emission
(Mulchaey, Wilson \& Tsvetanov 1996), and presents a steep continuum
with faint emission lines. We detected the \bhepg\ blend, \pabeta, and
a feature at \app\ 1.96\mic\ which we tentatively identify as
[\ion{Si}{vi}]\,1.962\mic. The lines are only barely spatially resolved
(Fig.~\ref{fig_n5_n13}), with \pabeta\ presenting some low surface
brightness emission up to \app\ 4\arcsec\ SE. The (\bhepg)/\pabeta\ 
ratio in the inner 4\arcsec\ ($|r| < $ 2\arcsec) is 1.31\mm 0.13.

{\it NGC\,1365:} The XD spectrum (Fig.~\ref{fig_specxd}a, bottom),
obtained along the axis of the ionization cone at PA = 130\degr, shows
a red continuum, with prominent permitted lines (\bhepd, \bhepg,
\pabeta); \palfa\ is detected above the atmospheric absorption. The
emission (Fig.~\ref{fig_n5_n13}) is resolved and similarly extended
symmetrically from the nucleus in all lines, with the exception of a
feature in \bhepg\ at \app\ 4\arcsec\ NW (the `anti-cone' direction)
which can be associated with an optical hot-spot in the circumnuclear
star-forming ring (Storchi-Bergmann \& Bonatto 1991; Kristen et al.
1997).The line ratios (\bhepg)/\pabeta\ and (\bhepd)/(\bhepg) are
essentially constant in the inner 4\arcsec, with values 2.27\mm 0.12
and 0.19\mm 0.02, respectively.

\begin{figure}
\psfig{file=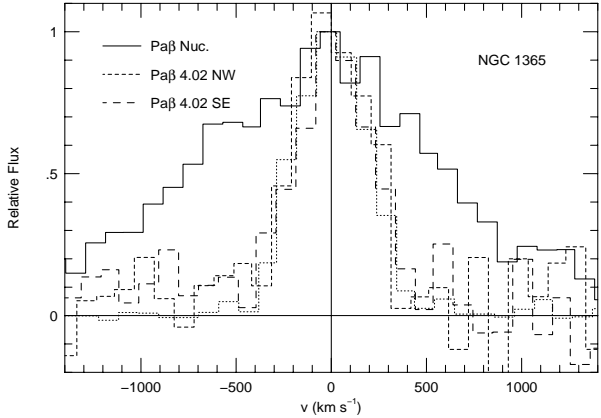,height=9cm,width=0.4\textwidth}
\caption{Comparison of the LR \pabeta\ spectral profiles of NGC\,1365:
nucleus (full line), 4\farcs02 NW (short dash), and 4\farcs02 SE (long
dash).  The dotted line is the \pabeta\ profile of NGC\,7009, which is
taken as representative of the instrumental resolution.}
\label{figlp_n1365}
\end{figure}

This galaxy was also observed with the LR grating in the J band. The
spectrum  (Fig.~\ref{fig_spechr}a, middle) shows very clearly the
presence of a broad component in the nuclear \pabeta\ profile
(Fig.~\ref{figlp_n1365}). The FWHM of the nuclear and extranuclear
profiles, obtained using multiple Gaussian component decomposition, is
listed in Table~\ref{tab_hrfw} for this and the other galaxies observed
with the LR and HR gratings. At 4\arcsec\ NW and SE, \pabeta\ is barely
resolved spectrally (see Table~\ref{tab_hrfw}).

The nuclear \pabeta\ profile can be described as the sum of a narrow
(N), broad (B) and very broad (VB) Gaussian components, with FWHM from
400 \kms\ to 2900 \kms. V\'eron et al. (1980) measured a flux of 7.7
\vzs\ \tento{-14}\ \ergcs\ in their broad (FWHM \app\ 1300 \kms)
component of \halfa. Associating this emission system with the FWHM
\app\ 1500 \kms\ component of \pabeta\ observed here, we obtain a broad
line region reddening of \av\ \app\ 7.7 mag for an intrinsic
\pabeta/\halfa\ ratio of 0.053. Since the broad lines of this galaxy
are known to be variable (Giannuzzo \& Stirpe 1996), not to mention
calibration differences, this value is to be regarded as only a first
order approximation to the actual broad-line region reddening.

Recently, Stevens, Forbes \& Norris (1999) suggested that the optical
broad lines in the nucleus of NGC\,1365 could originate not at the
Seyfert nucleus, but at one of the circumnuclear radio `hot spots'
(Morganti et al. 1999), located at \app\ 5\arcsec\ SW of the optical
nucleus,  as a result of a radio-supernova located there.  However, we
find this highly unlikely, since our spectrum, which was obtained with
a 1\arcsec\ slit, centred in the optical nucleus, and oriented at
almost right angles with the PA of the hot spot in question, would not
have included any of its emission.

\begin{figure}
\psfig{file=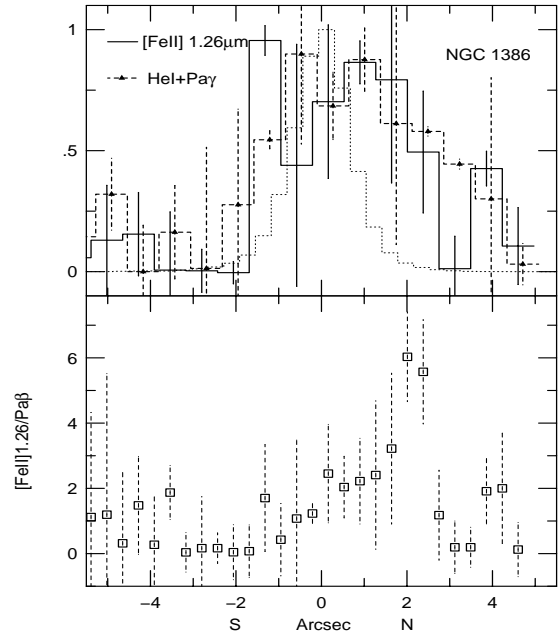,height=10cm,width=0.5\textwidth}
\caption{Spatial profiles and \bfeiia/\pabeta\ line ratio from the XD
spectrum of NGC\,1386. The dotted line is the stellar profile.}
\label{fig_n1386}
\end{figure}

{\it NGC 1386:} The XD spectrum (Fig.~\ref{fig_specxd}b, middle) was
obtained along the direction of the high excitation optical emission
line protrusion detected in the images of Storchi-Bergmann et al.
(1996). The brightest line is \bsiiia, with fainter \bhepg, \bfeiia,
and \pabeta. Although noisy, the resulting spatial profiles are
definitely extended along the [\ion{O}{iii}] emission direction
(Fig.~\ref{fig_n1386}). The \bhepg\ distribution is the most extended,
reaching at least  4\farcs5 N of the nucleus, the end of the slit. The
(\bhepg)/\pabeta\ ratio is 2.9\mm 1.0 in the inner 4\arcsec. The
\bsiiia\ and \bfeiia\ spatial profiles are resolved but more
concentrated, extending only up to 2\farcs5. In the inner 4\arcsec, the
\bsiiia/(\bhepg) ratio is essentially flat, with a value of 2.21\mm
0.21 , while \bfeiia/\pabeta\ increases from \app\ 0.9 at r =
2\arcsec\ S to \app\ 6 at 2\arcsec\ N, with a value of \dapp\ 2 in the
nucleus, indicating a higher excitation of the gas to the N, in
agreement with the optical images.

Several mechanisms for the origin of [\feii] emission in Seyferts
have been extensively discussed in previous works (Forbes \& Ward 1993;
Simpson et al. 1996a; Alonso-Herrero et al. 1997; Veilleux et al.
1997). In starbursts, if the mechanism generating the [\feii] emission
is directly related to the one producing the radio emission in (radio)
brigth supernova remnants, Colina (1993) calculated that
[\feii]\,1.26/\pabeta\ $\leq$ 0.4 is expected. The higher ratios
observed in Seyferts would then indicate ionization by X-rays from the
active nuclei or by shocks induced by the interaction of an outflow
with the surrounding medium. However, if the lifetime of the [\feii]
emission phase in single or multiple star formation bursts is longer
than in Colina's models, the [\feii]/\pabeta\ ratio can reach higher
values (Vanzi, Alonso-Herrero \& Rieke 1998), as the contribution from
the remnants becomes dominant over the gas ionization from the aging
\ion{H}{ii}\ regions.

Detailed stellar population analysis of NGC\,1386 (Schmitt,
Storchi-Bergmann \& Cid Fernandes 1999) does not indicate the presence
of a significant contribution from a young (10 Myr or less) component
in the inner 2\arcsec\ \vzs\ 2\arcsec, and therefore the
[\feii]/\pabeta\ values indicate that, as is also found in NGC\,2110
(Paper 1), a starburst related component of the [\feii] emission is not
important in this case. Since this object was not observed with the
higher resolution gratings, we do not have any information on the
[\feii] line profile, but the line broadening in \halfa\ and \niit 6584
within 1\farcs5 N of the nucleus detected by Weaver, Wilson \& Baldwin
(1991) raises the possibility of shocks as an ionization source for the
infrared [\feii] lines. Nagar et al. (1999) find the nuclear radio
source to be extended by 0\farcs4 in PA = 170\degr, a similar direction
to the extension of the optical emission lines.

\begin{figure}
\psfig{file=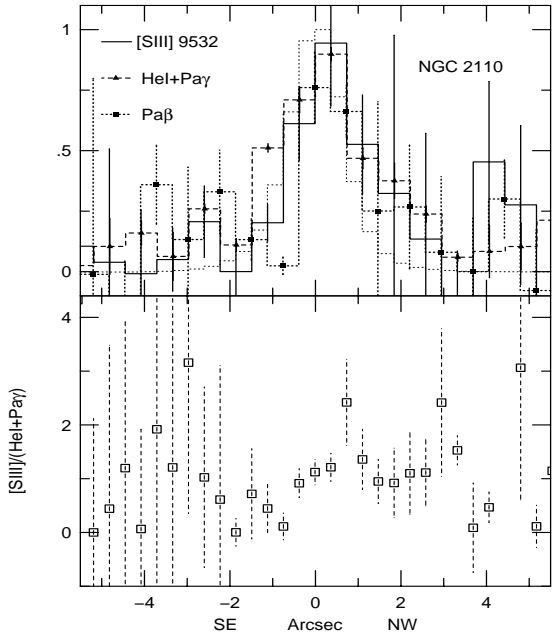,height=10cm,width=0.5\textwidth}
\caption{Spatial profiles and \bsiiia/\bhepg\ line ratio from the XD
spectrum of NGC\,2110. The error bars in the \bsiiia\ profile have been
excluded for clarity and are of similar amplitude to those in \pabeta.
The dotted line is the stellar profile.}
\label{fig_n2110}
\end{figure}

{\it NGC\,2110:} The HR/LR data for this galaxy were presented and
analysed in Paper I, and we found evidence of shocks and X-rays as
dominant sources of excitation for the [\feii] and \hh\ emission in the
inner few arcsec, respectively.  The XD spectrum shown in
Figure~\ref{fig_specxd}b (top panel) was obtained along the PA of the
radio axis, with \bsiiib, \bsiiia, \bhepg, \bfeiia, \pabeta, and
\bfeiib\  detected. The spatial profiles of \bsiiia, \bhepg\ and
\pabeta\ are clearly asymmetric (Figure~\ref{fig_n2110}), being
extended up to 3\arcsec\ towards the NW, the direction of the high
excitation optical emission. The [\ion{S}{iii}] and \pabeta\ spatial profiles
decline more quickly to the SE than the \bhepg\ profile, consistent
with the isophotes in the \halfa$+$[\ion{N}{ii}] image of Mulchaey et
al. (1996). The [\feii] spatial distribution (already discussed in Paper
I) is similar to that of \bsiiia.  The \bsiiia/(\bhepg) line ratio
rises from 0.44 at 2\arcsec\ SE to 1.0 at 2\arcsec\ NW. A similar
excitation gradient is seen in the [\ion{O}{iii}]\,\lb5007/\hbeta\ 
ratio (Wilson, Baldwin \& Ulvestad 1985). The \hh\ emission, clearly 
seen in the higher resolution spectra of Paper I, was not detected
here because of the low resolution of the XD spectra.

{\it NGC\,3281:} The XD spectrum is shown in Figure~\ref{fig_specxd}b
(bottom panel). It shows only faint [\ion{S}{iii}], \ion{He}{i}, and
\pabeta\ emission. The resulting spatial profiles were of too low S/N
ratio to allow any useful analysis of the extended emission.

\begin{figure}
\psfig{file=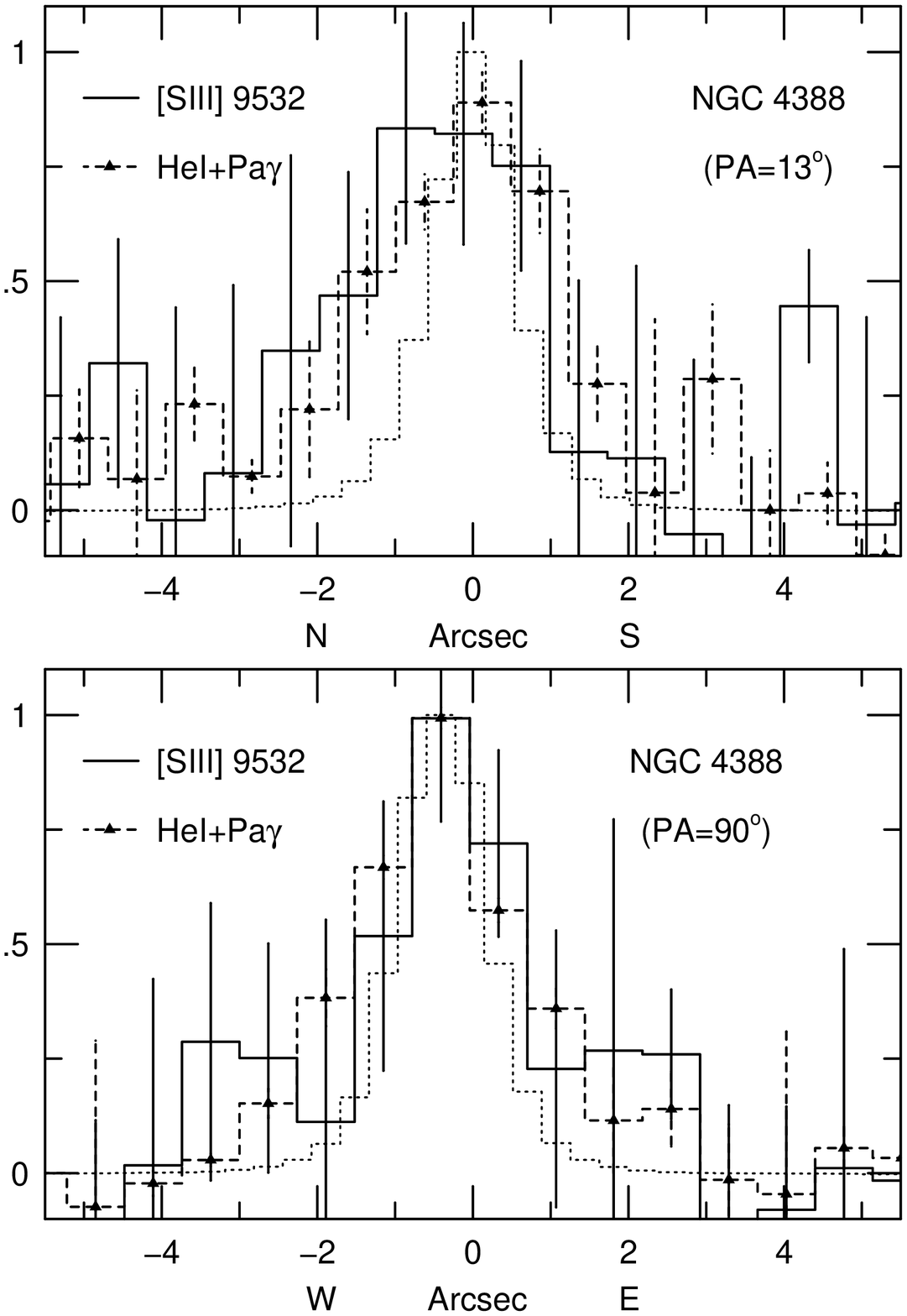,height=10cm,width=0.5\textwidth}
\caption{\bsiiia\ and \bhepg\ spatial profiles from the XD spectra 
of NGC\,4388, at PA = 13\degr\ (top) and PA = 90\degr\ (bottom).}
\label{fig_n4388n}
\end{figure}

{\it NGC\,4388:} Two sets of data were obtained for this object.  The
XD spectra (Fig.~\ref{fig_specxd}c, top and middle panels) were taken
approximately along and perpendicular to the axis of the optical
ionization cone, at PA=13\degr\ and 90\degr, respectively. The S/N is
low, with only \bsiiia\ and \bhepg\  detected.  The spatial profiles at
PA = 13\degr\ (Fig.~\ref{fig_n4388n}) are noisy, but more extended
towards the NE, the anti-cone direction. The narrow-band
\halfa$+$[\ion{N}{ii}] optical images of Veilleux et al.  (1999) show
an elongated structure in the inner 10\arcsec, which also appears to be
more extended to the NE direction (while the cone-like, kpc-scale high
excitation gas distribution is oriented towards the south). Along
PA=90\degr\ our emission line distributions are also resolved, but
essentially symmetric with respect to the centre. The line ratios are
almost constant in the inner 4\arcsec, with \bsiiia/(\bhepg) = 1.58\mm
0.84 (PA=13\degr) and 1.23\mm 0.74 (PA=90\degr). In this latter
spectrum, we also measured \bsiiia/\pabeta\ \app\ 4.30\mm 0.50 for this
same region.

Two HR spectra in the J band were obtained at the same position angles
as above (Figure~\ref{fig_spechr}a, bottom). The emission-lines
spatial profiles are only marginally resolved at both PAs, with
\bfeiia/\pabeta\ \app\ 0.7. Along the cone direction,
we clearly detect [\ion{S}{ix}]\,1.262\mic. 

For the PA = 90\degr\ spectrum, the IRS imaging mode was used to find
the position of the nucleus in the K band, which was found to be
located \app\ 1\arcsec\ N of the optical position. The slit was thus
centred offset by this amount from the optical nucleus. However, the
continuum is a factor of two lower than that of the PA =
13\degr\ spectrum, and the [\ion{S}{ix}] line is not as clearly
detected, suggesting that the true K-band nucleus was missed. The
spatial offset between the optical and infrared nuclei in NGC\,4388 has
been previously noted by Stone, Wilson \& Ward (1988), who found that
the 10 \mic\ emission peaks 'several arcsec' north of the optical
nucleus, rising the possibility of the mid-infrared source to be
associated with the radio ejecta, rather than with the nucleus proper.
      
\begin{figure}
\psfig{file=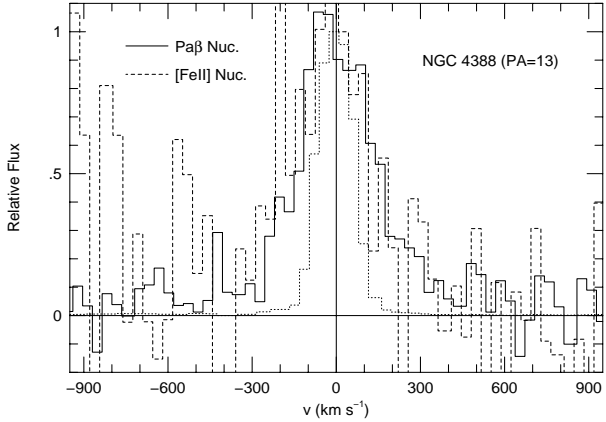,height=9cm,width=0.4\textwidth}
\caption{Comparison of the HR \pabeta\ (full line) and \bfeiia\ (dashed
line) nuclear spectral profiles of NGC\,4388 along PA=13\degr. The
dotted line is the instrumental profile, from the HeAr comparison lamp
spectrum.}
\label{figlp_n4388}
\end{figure}

\begin{figure}
\psfig{file=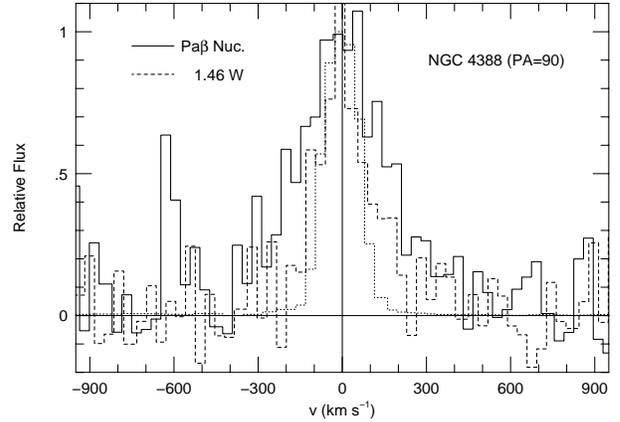,height=9cm,width=0.4\textwidth}
\caption{Comparison of the HR \pabeta\ spectral profiles of NGC\,4388
along PA=90\degr: nucleus (full line), and 1\farcs46 W (dashed line).
The dotted line is the instrumental profile, from the HeAr comparison
lamp spectrum.}
\label{figlp_n4388.90}
\end{figure}

The nuclear line profiles of \pabeta\ and \bfeiia\ from the
PA=13\degr\ spectrum are spectrally resolved, and very similar to each
other (Fig.~\ref{figlp_n4388}). Correcting the observed widths in
Table~\ref{tab_hrfw} by the instrumental broadening (adopted as the
FWHM of \pabeta\ in NGC\,7009) gives an intrinsic FWHM of \app\ 300
\kms\ for both lines. Along the perpendicular direction (PA = 90\degr),
the nuclear \pabeta\ profile is resolved and marginally broader than at
PA=13\degr, with a corrected FWHM of \app\ 378 \kms.  At 1\farcs5 W of
the nucleus the line is essentially unresolved (Fig.~\ref{figlp_n4388.90}).

{\it NGC\,5643:} The XD spectrum (Fig.~\ref{fig_specxd}c, bottom) was
obtained along the axis of the optical ionization cone, and shows
[\ion{S}{iii}], \bhen, \pabeta, and the tip of the \palfa\ emission
line. Only the \pabeta\ profile is marginally resolved spatially in the
central regions, but both this line and \bsiiia\ show a second emission
feature \app\ 3\arcsec\ E of the nucleus, which may be associated with
the filamentary extended emission seen in optical images (Schmitt,
Storchi-Bergmann \& Baldwin 1994; Simpson et al. 1997). In the central
2\arcsec, we measured \bsiiia/(\bhepg), \bsiiia/\pabeta, and
(\bhepg)/\pabeta\ ratios of 1.34\mm 0.62, 6.98\mm 1.03, and 5.18\mm
0.71, respectively.

\begin{figure}
\psfig{file=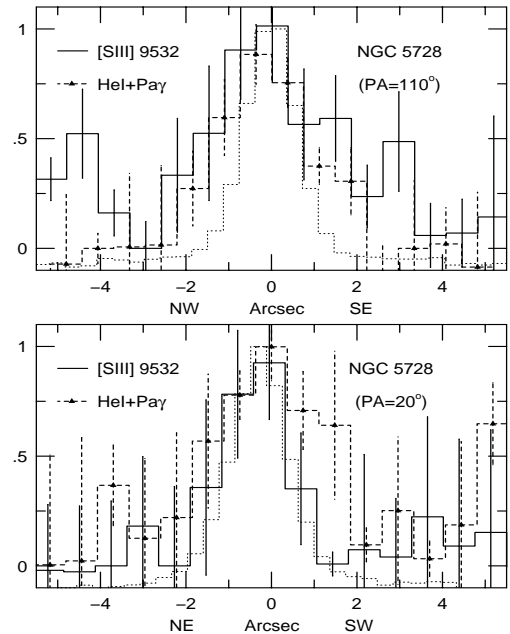,height=10cm,width=0.5\textwidth}
\caption{Spatial line profiles from the PA = 110\degr\ (top) and PA =
20\degr\ (bottom) XD spectra of NGC\,5728. The error bars in the
\bhepg\ profile have been omited for clarity and are of similar
amplitute to those in the \bsiiia\ distribution. The dotted line is the
stellar profile.}
\label{fig_n5728n}
\end{figure}

{\it NGC\,5728:} Two spectra (Fig.~\ref{fig_specxd}d) were obtained
with the XD grating, at PA = 110\degr, the optical ionization cone
direction, and perpendicular to it (PA = 20\degr). The S/N ratio is
low, and only \bsiiia, and \bhepg\ are detected. The spatial profiles,
shown in Figure~\ref{fig_n5728n}, are resolved along both directions,
although more extended at PA = 110\degr\ as expected from the optical
images of Wilson et al. (1993). However, while \bhepg\ is almost
symmetric, the \bsiiia\ emission is stronger to the NW, the opposite
direction of the optical cone.  The \bsiiia/(\bhepg) ratio in the inner
4\arcsec\ is 2.19\mm 0.85 along the cone, and 1.63\mm 0.94
perpendicular to it.

\begin{figure}
\psfig{file=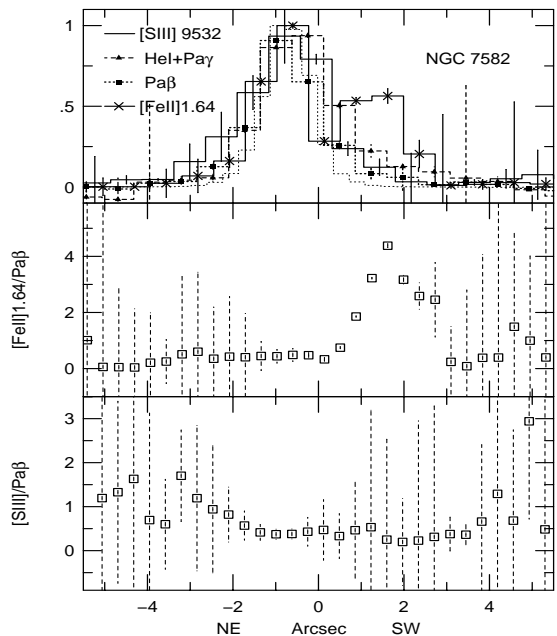,height=10cm,width=0.5\textwidth}
\caption{Spatial profiles and line ratios from the XD spectrum of
NGC\,7582.  The dotted line is the stellar profile. Note that the
origin of the spatial axis corresponds to the
nearby {\it continuum} peak for each line.}
\label{fig_n7582}
\end{figure}

{\it NGC\,7582:} This galaxy was observed with the slit at PA =
203\degr, which is within the optical ionization cone which has axis in PA
\dapp\ 250\degr) (Storchi-Bergmann \& Bonatto 1991).  The peaks of the
emission line spatial profiles (Fig.~\ref{fig_n7582}, top panel), are
offset from the peak of the nearby continua by 0\farcs7 -- 0\farcs8 to
the NE, the opposite direction to the ionization cone. All emission
lines, with the possible exception of \pabeta\ are extended towards the
cone. The \bfeiib\ emission presents a double peaked structure, with an
unresolved core plus a secondary bump reaching \app\ 2\farcs5 SW. The
\bsiiia\ profile is marginally extended to the NE. The double-peaked
structure of the \bfeiib\ emission is evident in the line-ratio
distribution: to the NE, the \bfeiib/\pabeta\ ratio
(Fig.~\ref{fig_n7582}, middle) is essentially constant at 0.45\mm 0.08,
reflecting the similar  light distribution of the two lines; to the SW,
the ratio increases sharply to 4.4 at 1\farcs6, and then falls back to
\app\ 0.3 at 2\farcs5.

In the absence of significant reddening, and taking the intrinsic ratio
between the [\feii] \lb\,1.64\mic\ and \lb\,1.26\mic\ lines as 0.75,
these values translate to a \bfeiia/\pabeta\ ratio between 0.6 and 6,
similar to those found above for NGC\,1386 or in Paper I for the
extended high excitation gas in NGC\,2110. Note that contrary to what
was observed for NGC\,2110, where the nuclear [\feii]/\pabeta\ ratio
was quite high (\app\ 7), the nucleus of NGC\,7582 presents a much
lower ratio than the extended emission, with only a small excess over
the expected ratio for young starbursts. Towards the cone, however, the
tenfold increase in the [\feii]/\pabeta\ ratio suggests the presence of
an additional source of ionization, either excitation by X-ray photons
from the active nucleus, escaping preferably towards the cone (due to
shadowing by the torus) or the interaction of the radio plasma with the
ambient medium. The recent radio maps of Morganti et al. (1999) show
that the 3.5 cm emission is indeed extended and spatially coincident
with the inner regions of the extended \halfa\ and [\ion{O}{iii}]
emission (Storchi-Bergmann \& Bonatto 1991).

If there is significant intervening reddening (\av\ \app\ 13 mag from
the [\feii] 1.64\,\mic/1.26\,\mic\ integrated line ratio from
Table~\ref{tab_xdflx}; or \av\ $\lesssim$\ 8 mag from the continuum
colours, see Section~\ref{sec_colours}), the above observed values
would translate to an intrinsic ratio \bfeiia/\pabeta\ $\gtrsim$\ 1 in the
cone region, still requiring  an additional source of ionization other
than young starbursts. A corresponding increase towards the SW is not
observed in the \bsiiia/\pabeta\ ratio (Fig.~\ref{fig_n7582}, bottom),
where we measured values of 0.61\mm 0.29 ($r < $ 3\arcsec\ NE) and
0.34\mm 0.13 ($r < $ 3\arcsec\ SW).

\begin{figure}
\psfig{file=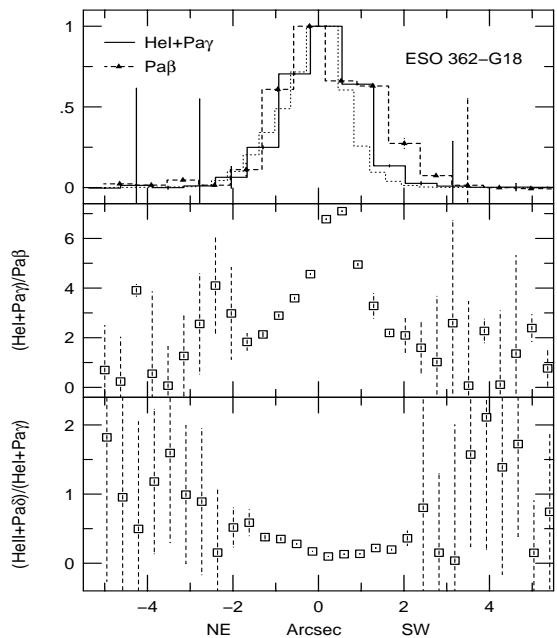,height=10cm,width=0.5\textwidth}
\caption{Same as Fig.~\ref{fig_n7582} for ESO\,362-G18.}
\label{fig_eso362}
\end{figure}

{\it ESO\,362-G18:}  Our XD spectrum of this Seyfert 1 galaxy
(Fig.~\ref{fig_specxd}e, top) was obtained along PA = 68\degr, which is
perpendicular to the direction of the axis of the approximately
cone-shaped high-excitation gas emission region (Mulchaey et al.
1996).  The spectrum presents bright broad lines including \bhepd,
\bhepg, \pabeta, and \palfa, which is located close to the atmospheric
absorption between the H and K bands and the flux must be regarded as a
lower limit. The permitted lines spatial profiles
(Fig.~\ref{fig_eso362}) are extended towards the SW, which can be
identified with the emission structure seen in the optical
\halfa$+$[\ion{N}{ii}] images of Mulchaey et al. (1996). The
distribution of the (\bhepg)/\pabeta\ line ratio is approximately
symmetric, with a maximum of \app\ 7.2 about 0\farcs5 SW of the
continuum peak, and decreasing to \app\ 1.5 at 3\arcsec\  both sides.
The behaviour of the (\bhepd)/(\bhepg) ratio is quite different, with
an  essentially constant value of 0.29 \mm 0.16 in the $r < $
2\arcsec\ region.

\begin{figure}
\psfig{file=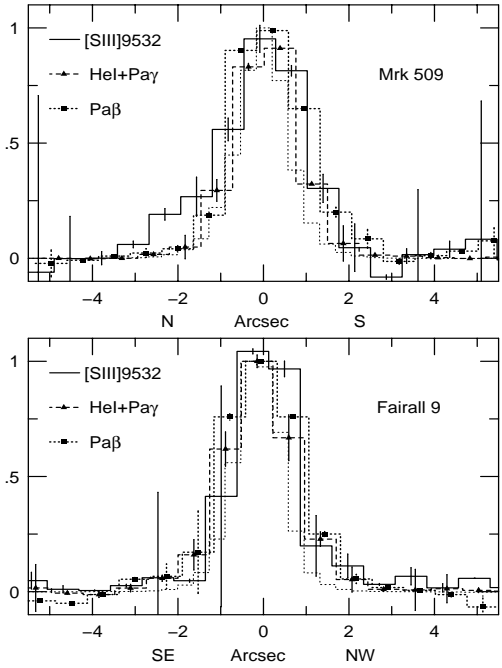,height=10.5cm,width=0.5\textwidth}
\caption{The spatial emission line profiles from the XD spectra of
Mrk\,509 (top) and Fairall\,9 (bottom). The dotted line is the stellar
profile.}
\label{fig_m5_f9}
\end{figure}

{\it Mrk\,509:} This is a Sy 1 galaxy suggested by Phillips et
al. (1983) as presenting a face-on outflow. The XD spectrum in
Figure~\ref{fig_specxd}e (middle) was obtained along the N--S
direction. The spatial profiles (Fig.~\ref{fig_m5_f9}, top) are
slightly resolved, and more extended towards the South, with the
exception of the \bsiiia\ line, which also shows extended emission up
to 3\arcsec\ N of the nucleus. The line ratios in the inner 4 --
5\arcsec\ are dominated by the nuclear source, with
\bsiiia/\pabeta = 0.42\mm 0.13, \bsiiia/(\bhepg) = 0.10\mm 0.03,
(\bhepg)/\pabeta\ = 3.50\mm 0.52, and (\bhepd)/(\bhepg) = 0.19\mm
0.01.

{\it Fairall\,9:} The XD spectrum of this classical Sy1 nucleus is very
similar to that of Mrk\,509 (Fig.~\ref{fig_specxd}e). The brightest
line is also \bhepg\ and the spatial profiles (Fig.~\ref{fig_m5_f9},
bottom) are essentially  unresolved, with \bsiiia/\pabeta,
\bsiiia/(\bhepg), (\bhepg)/\pabeta,  and (\bhepd)/(\bhepg) in the inner
4 -- 5\arcsec\ of 0.50\mm 0.08, 0.10\mm 0.03, 3.30\mm 0.33, and 0.16\mm
0.01, respectively.

{\it NGC\,1097:} We obtained a LR spectrum in the J-band of this LINER
nucleus in order to determine if the broad double-peaked profile
observed in \halfa\ (Storchi-Bergmann, Baldwin \& Wilson 1993) was also
present in \pabeta. The long-slit spectrum, however, shows \pabeta\
emission only from an \ion{H}{ii}\ in the circumnuclear star-forming
ring,  8\farcs7 SW of the nucleus (Fig.~\ref{fig_spechr}a, top panel).
This is a clear case of a very low luminosity, rather than dust
obscured active nucleus.

{\it IC\,5063:} Only a LR K-band spectrum (Fig.~\ref{fig_spechr}b, top) was
obtained, under non-photometric conditions. The orientation of the slit
was PA = 90\degr, which is close to that of the radio/optical/infrared
structures in the inner few arcsec (Simpson, Ward \& Kotilainen 1994;
Morganti, Oosterloo \& Tsvetanov 1998; Kulkarni et al. 1998), and both
\brgama\ and \hhl\ lines were detected.

\begin{figure}
\psfig{file=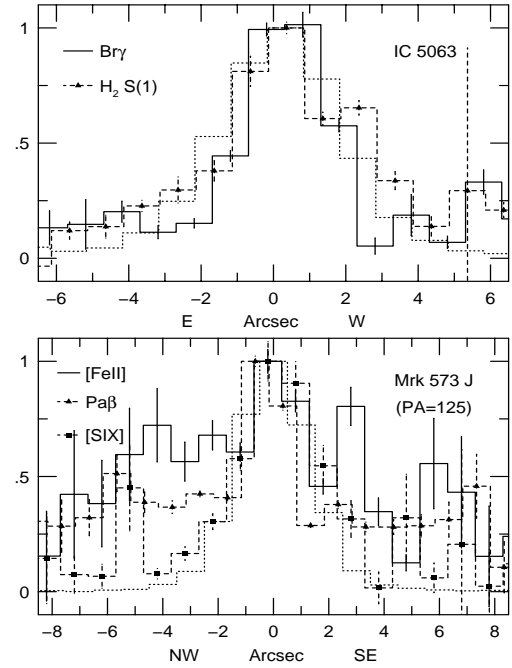,height=10.5cm,width=0.5\textwidth}
\caption{Spatial profiles from the LR spectra of IC\,5063 (K-band, top)
and  Mrk\,573 (J-band, bottom). The dotted line is a stellar profile.}
\label{figp_i5_m5}
\end{figure}

The \brgama\ spatial distribution is unresolved, while the \hh\ line
appears to be only slightly resolved (Fig.~\ref{figp_i5_m5}, top). The
\hh/\brgama\ ratio can be used to discriminate between the possible
excitation mechanisms for \hh\ emission: in star-forming regions, where
the main heating agent is the UV photons, \hh/\brgama\ $<$ 1.0 is
expected, while additional \hh\ emission excited by shocks or by X-rays
from the active nucleus increases the observed values to up to 3 or
more in Seyferts (Fischer et al. 1987; Moorwood \& Oliva 1990; Kawara,
Nishida \& Gregory 1990; Veilleux et al. 1997). In our spectrum, we
measured \hh/\brgama\ = 1.22\mm 0.20 in the inner 5\arcsec, which
constitutes marginal evidence of a contribution from the active
nucleus.The luminosity in the \hh\ line can be used to estimate the
mass of hot molecular hydrogen in the nuclear region. If the hot
\hh\ molecules are thermalized at T=2000 K, and assuming that the
emission in all \hh\ lines is 10 times that in the S(1) line (Scoville
et al. 1982, Veilleux et al. 1997), the observed \hh\ $\nu$=1$-$0 S(1)
luminosity of L(\hh) = 1.35 \vzs\ \tento{39}\ \ergs\ translates into a
hot \hh\ mass of about 450 \msol.

\begin{figure}
\psfig{file=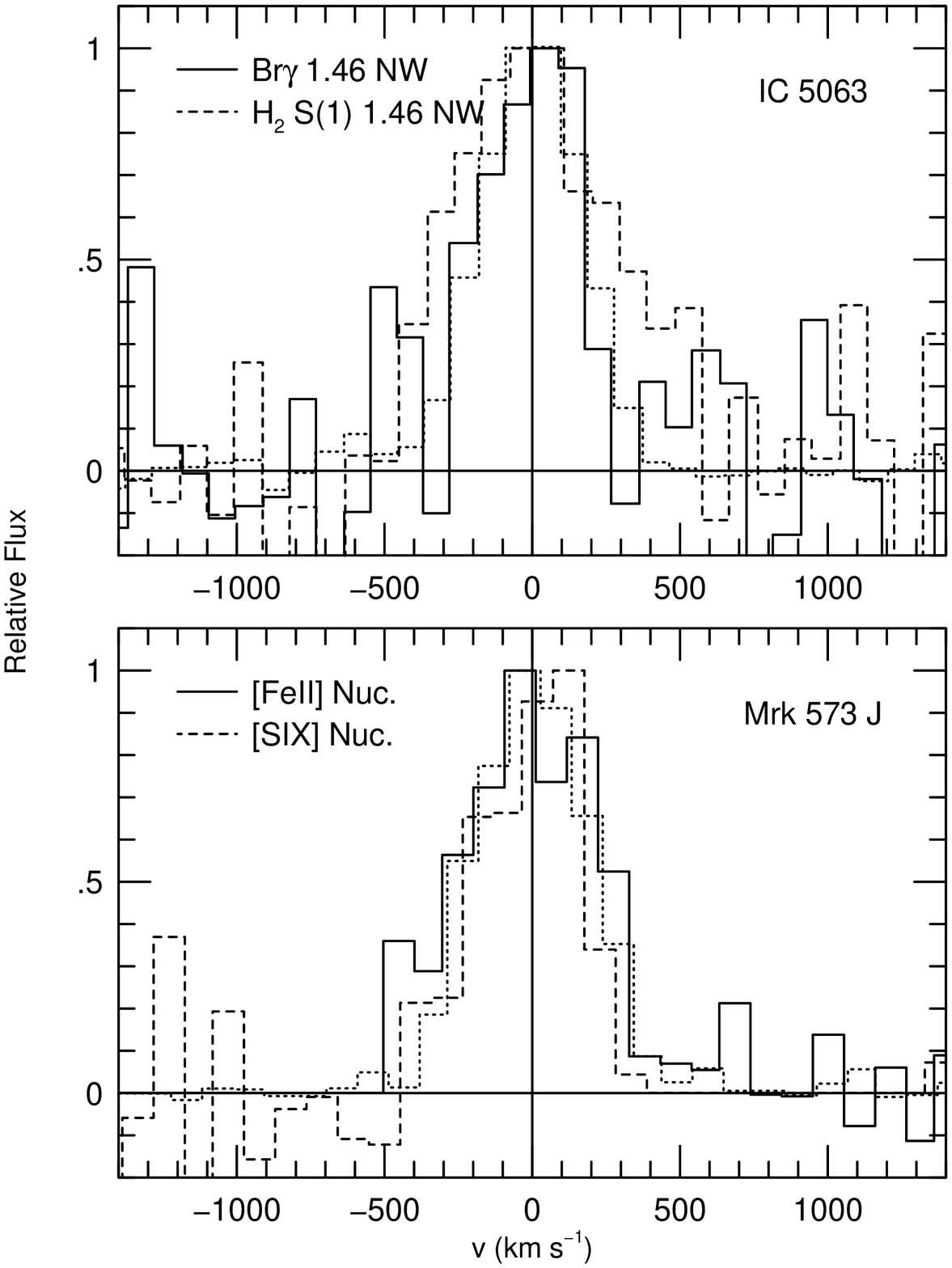,height=10.5cm,width=0.5\textwidth}
\caption{ Top: comparison of the LR spectral line profiles of the
K-band spectrum of IC\,5063: 1\farcs46 NW \brgama\ (full line) and
\hh\,S(1) (dashed line). Bottom: J-band (PA=125\degr) nuclear spectrum
Mrk\,573: \bfeiia\ (full line) and [\ion{S}{ix}] (dashed line)
profiles. The dotted lines in the top and bottom panels are,
respectively,  the \brgama\ and \pabeta\ profiles of NGC\,7009, which
are taken as representative of the instrumental resolution.}
\label{figlp_ic5_m5}
\end{figure}

The spectral profiles of the \hh\ and \brgama\ lines are unresolved,
except for the \hh\ emission in the 1\farcs5 NW spectrum
(Fig.~\ref{figlp_ic5_m5}, top), which has an observed FWHM of \app 630
\kms, corresponding to a intrinsic width of 484 \mm 100 \kms.

\begin{figure}
\psfig{file=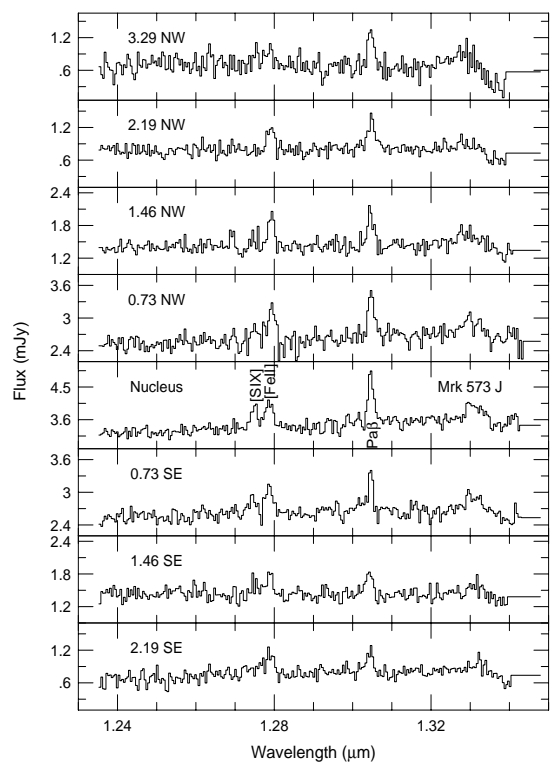,height=12cm,width=0.5\textwidth}
\caption{Individual J band spectra for Mrk\,573. Bin size is 0\farcs73
(1\farcs46 for the outermost position), centred at the distance from
the nucleus indicated in the individual panels.}
\label{figs_mrk573j}
\end{figure}

{\it Mrk\,573:} Two LR spectra were obtained for this galaxy. The
J-band spectrum (Fig.~\ref{fig_spechr}b, middle) was taken along the
radio continuum and optical emission-line axis at PA = 125\degr (Pogge
\& de Robertis 1995; Falcke, Wilson \& Simpson 1998), and presents
[\ion{S}{ix}], \bfeiia, and \pabeta\ emission. Individual spectra,
extracted in 0\farcs73 increments (Fig.~\ref{figs_mrk573j}), as well as
the spatial profiles (Fig.~\ref{figp_i5_m5}, bottom) clearly show that
the [\ion{S}{ix}] distribution is unresolved, while \pabeta\ and
\bfeiia\ are extended. The \bfeiia\ spatial profile presents a weak
central peak with a flat extension towards the NW, reaching up to
\app\ 8\arcsec, while the SE side is less prominent, but equally
extended. On the other hand, the \pabeta\ distribution is characterized
by a well-defined, spatially unresolved, central core, and extended
emission of almost equal intensity on both sides of the nucleus. The
\bfeiia/\pabeta\ ratio has a mean value of 0.77\mm 0.14 in the inner
5\arcsec.

\begin{figure}
\psfig{file=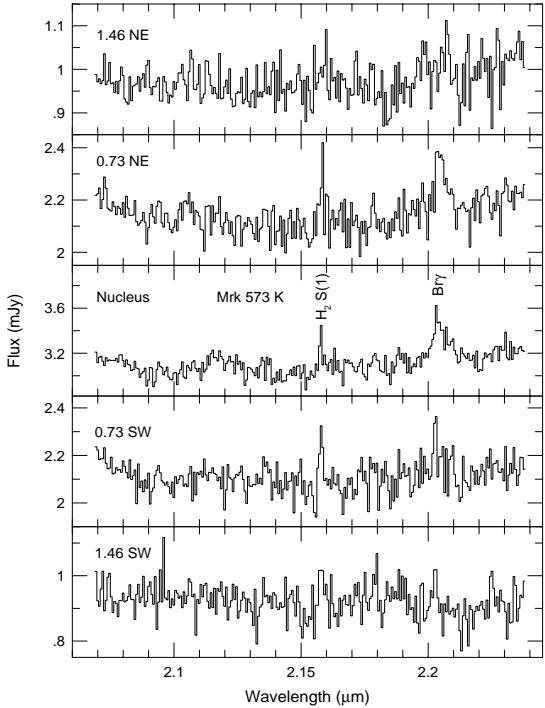,height=11.25cm,width=0.5\textwidth}
\caption{Same as Fig.~\ref{figs_mrk573j}, for the LR K band spectrum of Mrk\,573.}
\label{figs_mrk573k}
\end{figure}

A K band LR spectrum of Mrk\,573 (Fig.~\ref{fig_spechr}b, bottom) was
obtained at PA = 35\degr, perpendicular to the direction of the  radio
jet. \brgama, possibly broad, and narrow \hhl\ were
detected. Extracted spectra, in increments of 0\farcs73, are shown in
Figure~\ref{figs_mrk573k}.

\begin{figure}
\psfig{file=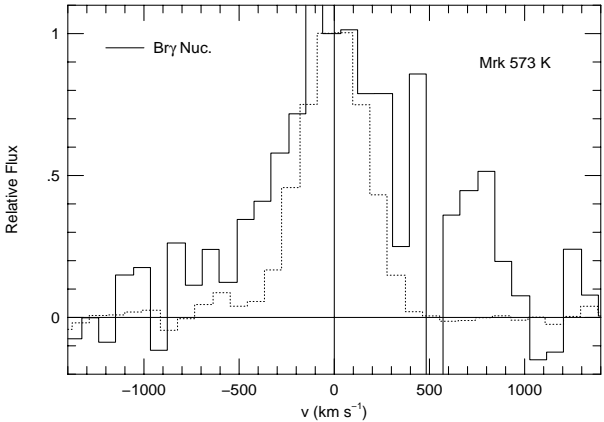,height=9cm,width=0.4\textwidth}
\caption{The nuclear \brgama\ LR spectral line profiles of Mrk\,573 (K
band, PA=35\degr). The dotted line is the \brgama\ profile of
NGC\,7009, which is taken as representative of the instrumental
resolution.}
\label{figlp_mrk573k}
\end{figure}

In the J band spectra, along PA = 125\degr, the spectral profile of
\pabeta\ is unresolved up to 3\farcs3 NW and SE.  The [\ion{S}{ix}]
line, detected only in the nuclear spectrum, is also unresolved, while
\bfeiia\ is marginally resolved spectrally in the nucleus but not
outside (Fig.~\ref{figlp_ic5_m5}, bottom panel). In the K band, the
\hh\ nuclear spectral profile is too noisy to yield any information,
while \brgama\ is clearly resolved (Fig.~\ref{figlp_mrk573k}), with an
intrinsic width of \app\ 590 \kms, and  some suggestion of a broad
component (FWZI \dapp\ 1900 \kms), which contains about 70 percent of
the total flux in the line, and is not observed in \pabeta\ (FWZI
$\lesssim$\ 1100 \kms).  Taking the upper limit for the flux of a
similar broad component in \pabeta\ as \app\ 5
\vzs\ \tento{-15}\ \ergcs\ (calculated from the rms of the continuum
around \pabeta, and a FWHM of 1300 \kms\ obtained from Gaussian
decomposition of the \brgama\ profile), we have
\brgama/\pabeta\ $\gtrsim$ 0.6, implying a reddening of
\av\ $\gtrsim$\ 9 mag in our line of sight to the broad line region in
this galaxy.

Except for the presence of the broad base in \brgama, our results are
in good agreement with those of Veilleux et al. (1997), which also
found that \pabeta\ and \hhl\ were narrower than \bfeiia\ or \brgama,
with this last line presenting an intrinsic width of \app\ 550 \kms.
Taking the narrow component of \brgama\ as containing about one third
of the observed line flux, the \hh/\brgama\ ratio is \app\ 1.2, and
would indicate some marginal evidence of \hh\ excitation by X-rays from
the active nucleus.  Using the same assumptions as for IC\,5063 the
observed luminosity of \hhl\ in the inner 3\arcsec\ translates into a
mass of hot \hh\ of \dapp\ 300 \msol.

\subsection{Continuum} \label{sec_colours}

In this section, we present and analyse \jmh\ and \hmk\ colour profiles
derived from our XD spectra.  Previous studies of the central (less
than 1 kpc) region of Seyfert galaxies have shown that their near-IR
colours are well explained either by a mixture of an evolved stellar
population plus emission from hot (800 -- 1200 K) dust, with varying
foreground reddening (Alonso-Herrero, Ward \& Kotilainen 1996,
hereafter AH96), or by a model more closely associated to the dusty
torus scenario, in which the starlight is not significantly reddened,
while the nuclear continuum source is represented by hot dust (T
\dapp\ 1200 K) reddened by \av\ \dapp\ 5 -- 30 mag (Alonso-Herrero et
al. 1998, hereafter AH98). The off-nuclear  colours show little
evidence of hot dust emission, except for a few cases where the effect
is attributed to the presence of massive stars acting as heating
sources.

To obtain the colour profiles, the calibrated spectra were rebinned to
0.8 arcsec/pixel, then multiplied by the transmission curve of the
corresponding filter, obtained from the CTIO web site\,\footnote{We
have used the curves of the filters in use with the CIRIM camera --
j40, h44, k50, for J, H, and K, respectively}, integrated along the
wavelength direction and converted from flux to magnitudes. With an
average seeing of 1 -- 1\farcs5, each profile is sampled by 1 -- 2
pixels.

\begin{figure}
\psfig{file=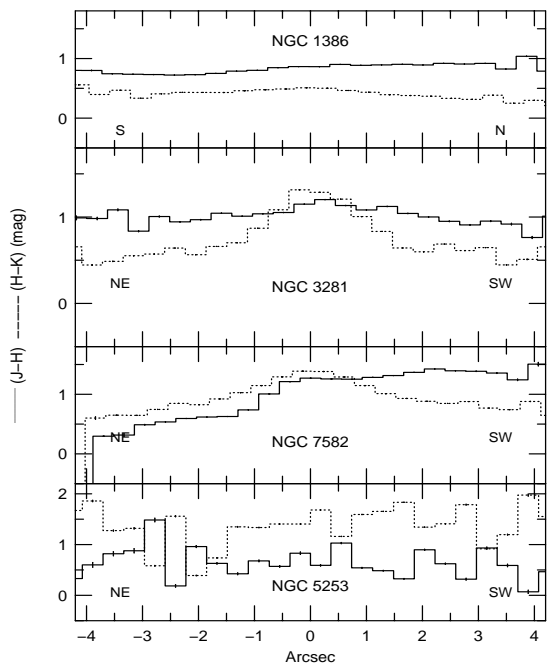,height=10.5cm,width=0.5\textwidth}
\caption{Four representative \jmh\ (full line) and \hmk\ (dotted line)
colour profiles derived from the XD spectra of the sample galaxies. The
orientation is the same as for the line profiles shown in
section~\ref{sec_profiles} and is also indicated in the individual
panels.} 
\label{fig_perfis2}
\end{figure}

Table~\ref{tab_xdclrs} lists the nuclear ($r <$ 0\farcs8) and `bulge'
(0\farcs8 $< r < $ 2\arcsec) colours derived from the above profiles.
These intervals correspond to linear scales of \app\ 45 to 750 pc for
the inner and outer radii in the Seyfert 2 galaxies in the sample. The
errors correspond to the rms around the mean of the colour in the
aperture, not the observational error. Figure~\ref{fig_perfis2} shows
four representative colour profiles: some are very flat, as in
NGC\,1386, while others show redder \hmk\ colour in the nuclear region
or asymmetric profiles, as in NGC\,3281 and NGC\,7582, respectively.
The colours in NGC\,5253 are highly inhomogeneous, as might be expected
from the clumpy nature of the distribution of the gas and presumably
the reddening in this galaxy.

The data in Table~\ref{tab_xdclrs} are in agreement with previous
conclusions (AH96, AH98, Kotilainen et al. 1992, hereafter K92) that
the nuclei of Seyferts tend to be redder than their bulges in one or
both \jmh, \hmk\ colours. Using the full profile, our data agrees with
the 1\farcs5 and 3\arcsec\ aperture photometry colours published in the
literature within 0.1 -- 0.2 mag for most objects, except NGC\,4388,
where our \hmk\ colour is \app\ 0.5 mag redder than the published
values (AH96, AH98). We also find that for a few galaxies there is a
0.3 -- 0.4 mag difference in the colours, most remarkably \jmh, of the
`bulge' regions at opposite sides of the nucleus.

The \jmh\ versus \hmk\ colour-colour diagrams of Figures~\ref{fig_bbm}a
to \ref{fig_bbm}f are based on the work of Alonso-Herrero et al.
(1998). The shaded square to the left represents the colours of bulges
of normal spiral galaxies (a late-type stellar population), while the
one to the right represents the typical colours of quasars, corrected
to zero redshift. The area bounded by dashed lines in the lower left
corner of Figs.~\ref{fig_bbm}c and \ref{fig_bbm}g corresponds to the
region occupied by a young/intermediate age stellar population
component with ages ranging from \tento{6} to \tento{8} years,
calculated from the models of Leitherer \& Heckman (1995). The arrow
in the upper-left hand corner represents the effects of 2 mag of visual
extinction. The `mixing curves' represented by the dashed lines are the
colours of the sum of a late-type stellar population and either a
black-body component (assumed to be representative of hot dust
emission) with temperatures $T = $ 600, 800, 1000, and 1200K, or a
nebular component, corresponding to hydrogen and helium continuum and
line emission. The dotted lines (diagonal slashes) are the loci of the
colours for the given ratio of dust (nebular) to stellar luminosities
in the K band. The results for each individual galaxy are described
below.

\begin{table*}
\begin{minipage}{150mm}
\caption{Nuclear and bulge colours  \label{tab_xdclrs}}
\begin{tabular}{lcccccccc}
\hline\hline
Object & \multicolumn{2}{c}{Nucleus (R $<$ 0\farcs8)} &  \multicolumn{6}{c}{Bulge (0\farcs8 $<$ R $<$ 2\arcsec)}\\
           & \jmh    & \hmk    &  & \jmh  &  \hmk  & & \jmh  &  \hmk \\
\hline
NGC\,526A       & 1.27\mm 0.03 & 1.24\mm 0.04 & NW & 0.88\mm
0.07 & 1.25\mm 0.10 & SE & 1.12\mm 0.03 & 0.99\mm 0.13 \\
NGC\,1365       & 1.07\mm 0.03 & 0.97\mm 0.03 & NW & 0.78\mm
0.03 & 0.94\mm 0.08 & SE & 1.05\mm 0.03 & 0.57\mm 0.05 \\
NGC\,1386       & 0.87\mm 0.01 & 0.49\mm 0.01 & S  & 0.77\mm 
0.02 & 0.45\mm 0.01 & N  & 0.89\mm 0.01 & 0.40\mm 0.02 \\
NGC\,2110       & 1.01\mm 0.01 & 0.94\mm 0.02 & SE & 1.02\mm 
0.02 & 0.66\mm 0.07 & NW & 1.01\mm 0.04 & 0.68\mm 0.05 \\
NGC\,3281       & 1.13\mm 0.03 & 1.21\mm 0.06 & NE & 1.02\mm 
0.01 & 0.72\mm 0.06 & SW & 1.08\mm 0.02 & 0.79\mm 0.10 \\
NGC\,4388 (13)  & 0.89\mm 0.01 & 1.42\mm 0.03 & N  & 0.88\mm 
0.02 & 1.19\mm 0.04 & S  & 1.02\mm 0.07 & 1.16\mm 0.04 \\
NGC\.4388 (90)  & 1.03\mm 0.02 & 1.51\mm 0.02 & W  & 0.85\mm 
0.05 & 1.35\mm 0.04 & E  & 1.01\mm 0.01 & 1.28\mm 0.08 \\
NGC\,5643       & 0.93\mm 0.01 & 0.54\mm 0.01 & W  & 0.94\mm 
0.02 & 0.41\mm 0.03 & E  & 0.82\mm 0.02 & 0.42\mm 0.05 \\
NGC\,5728 (110) & 0.94\mm 0.02 & 0.45\mm 0.03 & NW & 0.84\mm 
0.01 & 0.53\mm 0.02 & SE & 0.94\mm 0.01 & 0.32\mm 0.01 \\
NGC\,5728 (20)  & 1.04\mm 0.03 & 0.39\mm 0.03 & NE & 0.87\mm 
0.05 & 0.51\mm 0.01 & SW & 1.12\mm 0.01 & 0.11\mm 0.05 \\
NGC\,7582       & 1.21\mm 0.04 & 1.33\mm 0.03 & NE & 0.70\mm 
0.06 & 0.99\mm 0.07 & SW & 1.31\mm 0.03 & 1.02\mm 0.06 \\
ESO\,362-G18    & 0.90\mm 0.04 & 0.69\mm 0.04 & NE & 0.84\mm 
0.03 & 0.38\mm 0.05 & SW & 0.65\mm 0.01 & 0.44\mm 0.03 \\
Fairall\,9      & 1.14\mm 0.03 & 1.11\mm 0.02 & SE & 1.19\mm 
0.02 & 1.08\mm 0.07 & NW & 0.77\mm 0.05 & 1.06\mm 0.03 \\
Mrk\,509        & 0.98\mm 0.05 & 1.24\mm 0.02 & N  & 1.11\mm 
0.07 & 1.05\mm 0.13 & S  & 0.59\mm 0.06 & 1.07\mm 0.07 \\
\hline
\end{tabular}
\end{minipage}
\end{table*}

{\it NGC\,5253:} Fig.~\ref{fig_bbm}c -- the data points are scattered
over the bottom right of the diagram. Since the long slit spectrum
indicates little or no continuum emission, the data can be understood
as nebular emission plus spatially inhomogeneous reddening and/or dust emission. From the \brgama/\pabeta\ ratio, we found regions
with \av\ \app\ 8 mag, which would be enough to shift the points from
the nebular emission curve to the observed location at the middle of
the diagram.

{\it NGC 526A:}  Fig~\ref{fig_bbm}b -- the data points follow different
paths in the cone (SE) and anti-cone (NW) direction, with the
\jmh\ colour in the region 0\farcs8 -- 1\farcs4 SE about 0.3 mag redder
than to the NW. This behaviour suggests difference in the underlying
stellar population content between the two sides of the nucleus. As can
be seen in Fig.~\ref{fig_bbm}, the loci of old (left grey square) and
young-intermediate (area at bottom left limited by dashed lines)
stellar populations are separated by \app\ 0.8 mag in \jmh, with very
little difference in the corresponding \hmk\ colours.  Reddening
effects are also very likely present and cannot be discarded as
contributing to the observed colours. In the central bin  (0\farcs8),
the colours suggest hot  dust emission dominates the continuum.

{\it NGC\,1365:} Fig~\ref{fig_bbm}f -- the colours behave somewhat
similarly to NGC\,526A, although with the nuclear colour corresponding
to a smaller ratio of  hot dust to starlight emission. The bluer
\jmh\ colour towards the NW may result from contamination of the
`normal', old  bulge population by younger stars in the `hot spots'
(see Storchi-Bergmann \& Bonatto 1991; Morganti et al. 1999).

{\it NGC\,1386:} Fig.~\ref{fig_bbm}b -- Both colour profiles
(Fig.~\ref{fig_perfis2}) are essentially flat, with \jmh\ rising
slightly towards the North (the optical cone direction) and
\hmk\ showing only a hint of a redder colour in the nuclear region. In
the colour-colour plot, the points concentrate near the pure bulge
region, with little or no contribution from hot dust. The orientation
of both the high excitation optical and radio emission
(Storchi-Bergmann et al. 1996; Weaver et al. 1991), as well as the fact
that NGC\,1386 contains water vapour megamaser emission (Braatz, Wilson
\& Henkel 1996), suggest that the collimating structure is seen edge-on.
Therefore, the hot dust in the inner regions may be obscured by colder
dust further out in the disk, resulting in a small contribution from
the nucleus to the near infrared continuum.

{\it NGC\,2110:} Fig.~\ref{fig_bbm}h -- The analysis of the J and K band
continua presented in Paper I indicates the presence of a hot dust
component in the inner \app\ 150pc, consistent with emission by a
circumnuclear torus. The observed colour profiles confirm this
finding:  while the \jmh\ distribution is essentially flat across the
nucleus, the \hmk\ colour increases by about 0.5 mag in the inner
2\arcsec\ relative to the external regions. In the colour-colour
diagram, ignoring the foreground galactic reddening
(\ebv$_G$\ \app\ 0.36; Burstein \& Heiles 1982), the points trace a
continuous path from pure bulge colours  to a combination of bulge and
a \app\ 1200 K black-body component at the nucleus.

{\it NGC\,3281:} Fig.\ref{fig_bbm}g -- The colour profiles
(Fig.~\ref{fig_perfis2}) present a similar behaviour to that observed in
NGC\,2110, with the variation in \hmk\ colour being even more
pronounced, with a 0.7 mag decrease from the nucleus to the
bulge. This would indicate a higher fraction of hot dust emission in the
nuclear region than in the NGC\,2110 case, but of a somewhat lower
temperature (\app\ 1000 K).

\begin{figure*}
\psfig{file=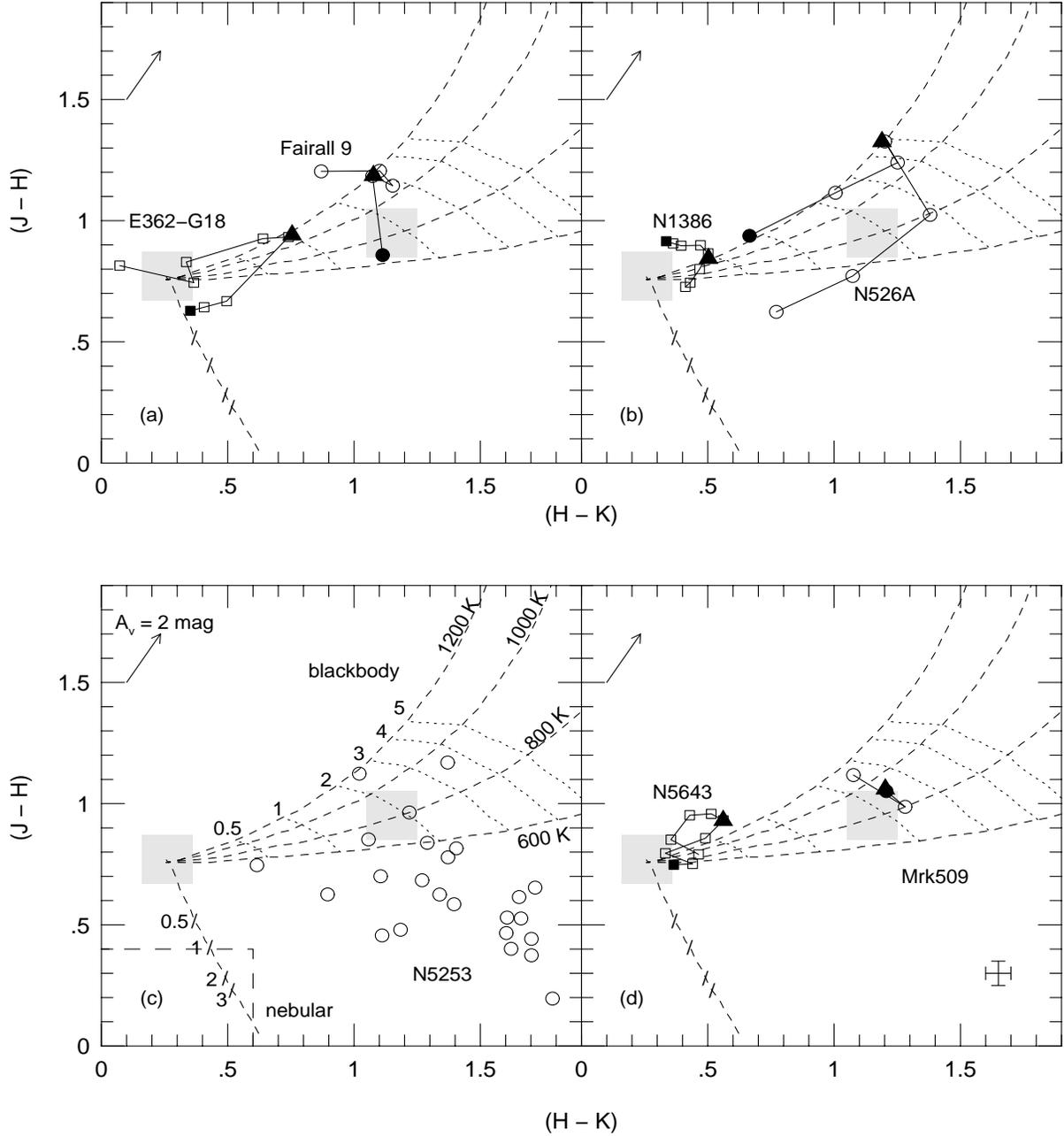,height=22cm,width=\textwidth}
\caption{\jmh\ vs \hmk\ colour-colour diagrams (adapted from
Alonso-Herrero et al 1998). See text for an explanation of the mixing
curves and other symbols. For each object, the filled triangle
represents the nucleus, while the filled circle/square is the most
distant point from the nucleus on the cone (positive) side of the
profile. The bin size is 0\farcs8 and the spatial interval (in arcsec)
represented is as follows: (a) ESO\,362-G18 (-3.4 NE,3 SW) and Fairall\,9
(-2.6 SE,1.4 NW); (b) NGC\,1386 (-3.4 S,3 N) and NGC\,526A (-2.6 NW,3
SE); (c) NGC\,5253 (-4 NE,4 SW; bin 0\farcs4); (d) NGC\,5643 (-3.4 W,3.8 E) and Mrk\,509 (-1.8 N,1.4 S).}
\label{fig_bbm}
\end{figure*} 

{\it NGC\,4388:} Fig.~\ref{fig_bbm}e -- The \hmk\ colour is redder at
the nucleus than in the bulge at PA = 13\degr, similar to what is
observed in NGC\,2110 and NGC\,3281. The effect is much less apparent
at PA = 90\degr, which can be understood if, as mentioned in
Section~\ref{sec_profiles}, we missed the nucleus on this
observation.   Both PA's occupy the same region in the colour diagram,
and the data points for the whole inner 5--6\arcsec\ (\app\ 1 kpc) are
shifted towards redder \hmk\ colours when compared with other published
values (K92, AH98).

{\it NGC\,5643:} Fig.~\ref{fig_bbm}d -- The colour profiles are flat
across the nucleus. In the colour diagram, the points cluster near
the bulge colours, with a small contribution from dust emission or a
foreground reddening of \av\ \app\ 1.5 --  2 mag. Again, this could be
a case of a torus either too cold to emit in the near infrared or with inner regions obscured from our line of sight by the outer ones.

{\it NGC\,5728:} Fig.~\ref{fig_bbm}e -- For PA = 110\degr, both colour
profiles are rather flat. In the colour-colour plot, the data points
cluster close to normal bulge colours, similar to what is observed
for NGC\,5643. Along the perpendicular direction (PA = 20\degr),
\jmh\ is somewhat redder and \hmk\ bluer towards the SW than to the
NE.

{\it NGC\,7582:} Fig.~\ref{fig_bbm}h -- The \hmk\ profile
(Fig.~\ref{fig_perfis2}) exhibits the same behaviour as observed in
NGC\,2110 and NGC\,3281, with the nuclear colour \app\ 0.5 mag redder
than the bulge. The \jmh\ profile, however, rises steadily to the SW,
with a 0.8 mag increase from 3\arcsec\ NE to 3\arcsec\ SW (the cone
side). In the colour-colour diagram, the nucleus corresponds to a
dominant contribution from \app\ 1000 K black-body emission. The data
points starting at 3\arcsec\ NE can be interpreted as a reddened
evolving starburst, with increasing reddening towards the nucleus (up
to \av\ \app\ 8 mag). The colours observed on the SW (cone) side of the
nucleus are suggestive of a reddened , old stellar population.

{\it ESO\,362-G18:}  Fig.~\ref{fig_bbm}a --  The data points start with
bulge colours at 3\arcsec\ NE, run along a line of increasing
contribution from hot dust, up to a maximum of equal contributions from
both sources at the nucleus, and then return with \jmh\ colours
\app\ 0.2 mag bluer towards the SW. This effect can again be explained
by a different underlying stellar population: an evolving starburst is
bluer in \jmh\ but occupies almost the same region in \hmk\ as the
typical late-type, bulge population.

{\it Mrk\,509} and {\it Fairall 9:}  Fig.~\ref{fig_bbm}d and
\ref{fig_bbm}a -- the PSF of the nucleus dominates the inner
4\arcsec\ due to the brightness of the central source. Since the
sources are unresolved, the data  points cluster around the colours
characteristic of quasars, as expected for these type 1 Seyfert
galaxies.

\begin{figure*}
\psfig{file=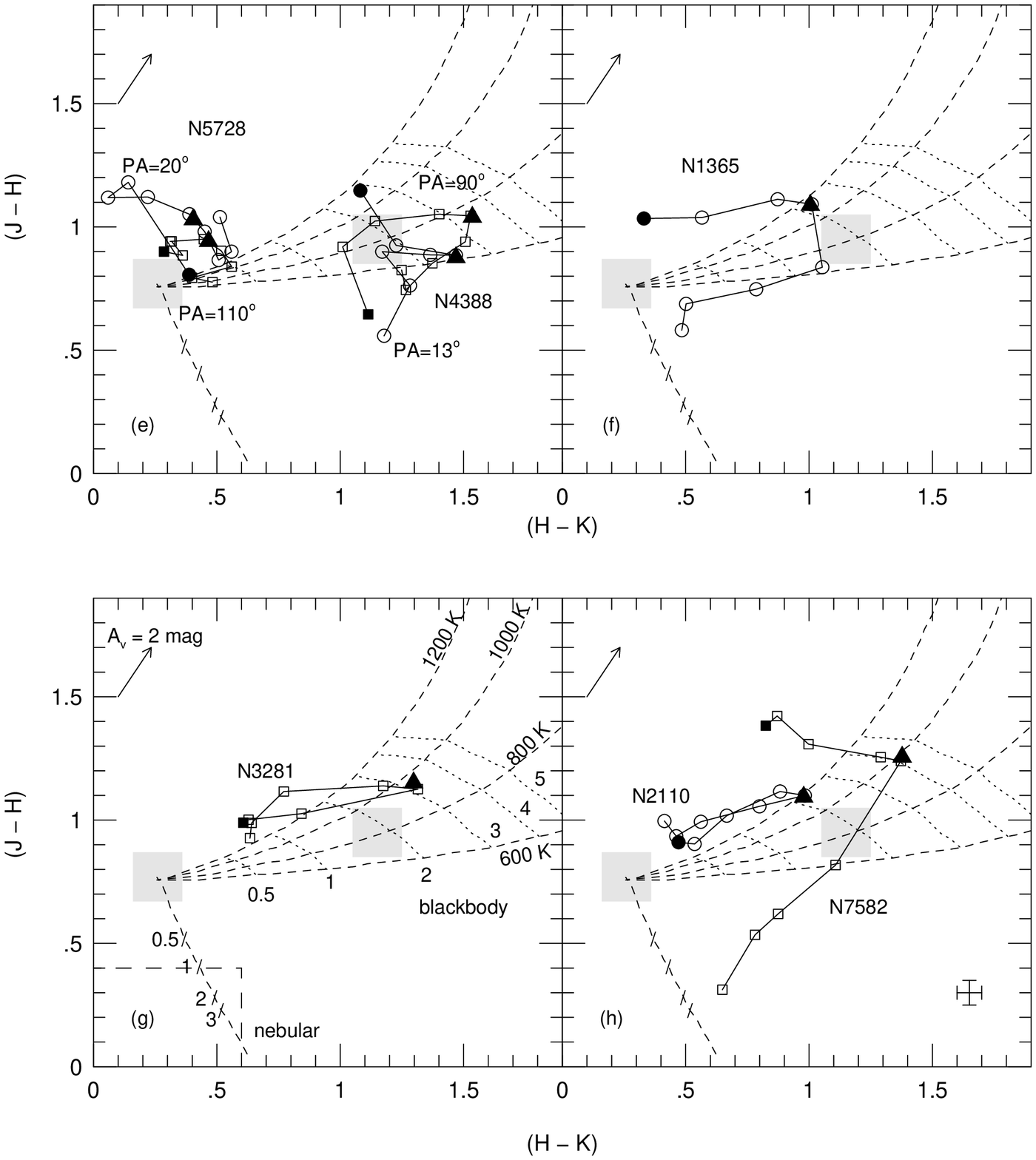,height=22cm,width=\textwidth} 
\contcaption{~(e)
NGC\,5728 (circles) PA=20\degr\ (-3.4 NE,3.8 SW), (squares)
PA=110\degr\ (-3.4 NW,3.8 SE) and NGC\,4388 (circles) PA=13\degr\ (-3.4
NE,2 SW), (squares) PA=90\degr\ (-3.4 W,3.8 E); (f) NGC\,1365 (-2.8 NW,3.8
SE); (g) NGC\,3281 (-3.4 NE,3 SW); (h) NGC\,2110 (-3.4 SE,3.8 NW) and
NGC\,7582 (-3.4 NE,3.8 SW).}
\end{figure*}

\section{Summary and conclusions} \label{sec_concl}

We obtained near-infrared, long-slit spectra for a sample of 12
Seyfert 2s, 3 Seyfert 1s, and the starburst galaxy NGC\,5253, using as
the main selection criterion the presence of extended high-excitation
emission observed in optical images. From our analysis of the spatial,
and spectral emission-line profiles, and of the continuum \jmh\ and
\hmk\ colours, we conclude that:

\begin{itemize}
\item consistent with the selection criteria, 9 of the 12 objects 
with  extended optical emission lines
also present extended emission lines in the near-IR along the position
angle of the optical ionization cones or radio emission. In three
cases, however, we found that the near-IR emission lines were also
extended along PAs oriented perpendicular to the optical/radio emission
axis. The clearest case where high-excitation near-IR line emission traces the
optical ionization structure is NGC\,7582, where the emission lines in
the inner few arcsec are extended along the direction of the optical
cone.  Interestingly, the maxima of the emission-line distributions in
this galaxy are consistently offset from the peak of the continuum
emission by \app\ 0\farcs8 in the direction opposite to the optical
cone.  Besides the central peak, the spatial profile of the
\bfeiib\ line in NGC\,7582 presents a secondary structure between
0\farcs5 and 2\farcs5 SW, with the \bfeiib/\pabeta\ ratio increasing
from \dapp\ 0.5 at the position of the continuum peak to 4.4 at
1\farcs6 SW, indicating a different ionization mechanism for the
[\feii] line in this region. A similar result was found for NGC\,1386,
for which the \bfeiia/\pabeta\ ratio rises from \dapp\ 2 to 6 between
the nucleus and the optical extended emission-line region at
2\arcsec\ N.  These high values and the presence of radio emission
and/or a high velocity nuclear outflow spatially coincident with the
optical ENLR suggest that shocks may play a role as an excitation
mechanism for the off-nuclear [\feii] line in both galaxies.

\item the low resolution (XD) spectra are not sufficiently sensitive to detect
the expected \hh\ emission from the torus. However, in the few cases
for which we have a high resolution K band spectra (Mrk\,573, IC\,5063,
NGC\,2110), the \hh\ S(1)\,\lb 2.12\mic\ was clearly detected.

\item broad components have been observed in the permitted emission
lines of the Seyfert 1 galaxies, in nuclear \pabeta\ in NGC\,1365,
and very likely in \brgama\ in Mrk\,573. In NGC\,1365 the flux of the
broad (FWHM \dapp\ 1500 \kms) component of \pabeta\ was used to
estimate a reddening of \av\ \app\ 8 mag towards the BLR. A lower limit
of \app\ 9 mag was obtained for \av\ towards the BLR of Mrk\,573. In
NGC\,4388 and Mrk\,573 the [\ion{S}{ix}]\,\lb 1.262\mic\ emission line
was detected, and found to be spatially unresolved in both objects.
With an ionization potential of 0.328 keV, the detection of this
species is an indirect evidence of significant soft X-rays continuum
flux in these galaxies. In NGC\,5253, the \brgama/\pabeta\ ratio was
used to map strong reddening variations, from \av\ \app\ 0 to 6 mag
within the central 8\arcsec.

\item spatial \jmh\ and \hmk\ colour profiles were derived from our
spectra.  For most Seyfert 2s, the nuclear colours are redder than the
extranuclear values, with the nuclear continuum being dominated by hot
(T \app\ 1000 K) dust emission in NGC\,1365, NGC\,2110, NGC\,3281,
NGC\,7582, and ESO\,362-G18. In NGC\,1386, NGC\,5643, and NGC\,5728,
the nuclear colours are consistent with a continuum dominated by the
emission from the underlying stellar population plus a smaller
contribution from dust and/or foreground reddening effects. The Seyfert 1 nuclei, Mrk\,509 and Fairall\,9, have colours similar to quasars, as expected.

\item the  galaxies NGC\,526A, NGC\,1365, NGC\,7582, and ESO\,362-G18
show a clear difference between the \jmh\ colours on each side of the
nucleus. Since late-type (bulge) and young/intermediate stellar
populations differ mostly in their \jmh\ colour, \hmk\ being
approximately the same, we explain this effect by differences in the
underlying stellar population, very likely coupled with reddening.  The
observed colour differences could indicate the presence of a
younger/intermediate age component, associated, for example, with the
star-forming rings known to exist in some of the observed galaxies, and
which could provide a larger contribution to the near-IR continuum on
one side of the nucleus relative to the other.

\end{itemize}

\section*{Acknowledgements}

This work has made use of NASA's Astrophysics Data System Abstract
Service (ADS), and of the NASA/IPAC Extragalactic Database (NED) which is
operated by the Jet Propulsion Laboratory, California Institute of
Technology, under contract with the National Aeronautics and Space
Administration. Research partially supported by the Brazilian Agencies
FAPERGS and CNPq.



\bsp

\label{lastpage}

\end{document}